\documentclass[11pt,a4paper]{article} 

\usepackage{amsmath}  
\usepackage{amssymb}  
\usepackage{amsthm}
\usepackage{mathabx}   
\usepackage{amsfonts} 
\usepackage{mathrsfs}  
\usepackage{bm}             
\usepackage{hyperref} 
\usepackage{graphicx} 
\usepackage{color,xcolor}
\usepackage{tabularx}
\usepackage[title]{appendix}
\usepackage{titlesec}
\titlelabel{\thetitle.\quad}

\usepackage{geometry}	
\usepackage{indentfirst} 
\numberwithin{equation}{section}   

\theoremstyle{plain}
\newtheorem{theorem}{Theorem}[section]
\newtheorem{theorem*}{Theorem}

\theoremstyle{definition}

\newtheorem{remark}[theorem]{Remark}

\DeclareMathOperator{\tr}{tr}
\DeclareMathOperator{\Div}{Div}

\begin{document}
	
\title{\bf A refined dynamic finite-strain shell theory for incompressible hyperelastic materials: equations and two-dimensional shell virtual work principle}
\author{Xiang Yu$^1$, Yibin Fu$^2$, Hui-Hui Dai$^{1,*}$ }
\date{$^1$Department of Mathematics, City University of Hong Kong, Kowloon Tong, Hong Kong\\
$^2$School of Computing and Mathematics, Keele University, Keele, UK}
\maketitle

\begin{abstract}
	Based on previous work for the static problem, in this paper we first derive one form of dynamic finite-strain shell equations for incompressible hyperelastic materials that involve three shell constitutive relations. In order to single out the bending effect as well as to reduce the number of shell constitutive relations, a further refinement is performed, which leads to a refined dynamic finite-strain shell theory with only two shell constitutive relations (deducible from the given three-dimensional (3D) strain energy function) and some new insights are also deduced. By using the weak formulation of the shell equations and the variation of the 3D Lagrange functional, boundary conditions and the two-dimensional (2D) shell virtual work principle are derived. As a benchmark problem, we consider the extension and inflation of an arterial segment. The good agreement between the asymptotic solution based on the shell equations and that from the 3D exact one gives verification of the former. The refined shell theory is also applied to study the plane-strain vibrations of a pressurized artery, and the effects of the axial pre-stretch, pressure and fibre angle on the vibration frequencies are investigated in detail. 
\end{abstract}

\footnotetext{$^*$Corresponding author}
\footnotetext{\textit{Email address}: \texttt{mahhdai@cityu.edu.hk}}
\footnotetext{The work described in this paper is fully supported by a GRF grant (Project No.: CityU 11303718) from the Research Grants Council of the Government of HKSAR, P.R. China.}

\section{Introduction}

In recent years, biological materials have attracted a lot of interest; see, for example, the review article by Holzapfel and Ogden \cite{holzapfel2010constitutive} on constitutive modelling of arteries. There are two noteworthy properties of biological materials. 
\noindent materials.  One is they are very soft and can undergo large elastic deformations with finite-strain; the other is that the volume is preserved during the deformation. So, they are normally modelled as incompressible hyperelastic materials. Many biological tissues and organs are thin structures. Due to the complexity of the 3D formulation and the cost and ineffectiveness of 3D computations (in particular, for post-bifurcation solutions), often one needs to use a 2D shell model to study their behaviors.

Shell theories have a long history, which date back to the pioneering work of Love \cite{love1888xvi} in 1888. Since then, they have been studied extensively during the past 130 years.  Numerous works on shell theories have been done in the framework of linearized elasticity  and/or linear constitutive relation with geometric nonlinearity. Here, the focus is on soft materials modelled by a strain energy function with incompressibility constraint, for which one needs to consider material nonlinearity. It is out of the scope of the present study to give an extensive review on linear shell theories or those with geometric nonlinearity, and for a selected review, we refer to Li {\it et al.} \cite{li2019consistent}.  Instead, we only give a review on {\it derived} shell theories for {\it incompressible hyperelastic materials}, for which, relatively speaking, there are not so many works. 

In \cite{makowski1986finite}, Makowski and Stumpf formulated a finite-strain shell theory for incompressible hyperelastic materials by assuming the material lines normal to the shell surface remain straight during the deformation. Itskov \cite{itskov2001generalized} assumed that the position vector in the deformed shell is linear in the thickness variable (with six parameters). The incompressibility constraint is used to eliminate the transverse normal strain, and based on which, a numerical shell theory with five parameters for a generalized orthotropic incompressible hyperelastic material was developed.  In \cite{chapelle2004asymptotic}, Chapelle {\it et al.} examined whether the plane stress assumption or the asymptotic limits of thickness can commute with the incompressibility constraint, justifying the usages of classical shell models and a modified 3D shell model in the incompressible conditions. In Kiendl  {\it et al.} \cite{kiendl2015isogeometric}, a shell theory for compressible and incompressible isotropic hyperelastic materials was developed based on the Kirchhoff-Love kinematics  which includes the assumptions of zero transverse normal stress and straight and normal cross-sections, and then an isogeometric discretization was introduced for numerical computation. Recently, Amabili {\it et al.} \cite{amabili2019nonlinear}, for a tube (a special kind of shells), developed a shell theory for incompressible biological hyperelastic materials by assuming the in-plane displacement components are third-order polynomials of the thickness variable while the out-plane component is a fourth-order polynomial. Further simplification in that work includes the dropping of certain nonlinear terms in the strain-displacement relations and incompressibility condition, which enables one to represent the four coefficients in the out-plane displacement in terms of other unknowns. As a result, a nine-parameter shell theory was obtained.  All the works mentioned above employ {\it ad hoc} assumptions and cross-thickness integrations to eliminate the thickness variable. As a result, one cannot expect that the resulting shell theories are consistent with the 3D field equations, top and bottom traction conditions and incompressibility condition in a pointwise manner. It is difficult to assess the reliability of such inconsistency for general loading. Also, when higher-order expansions are used, higher-order resultants need to be introduced but their physical meanings are not clear.  Thus, it is more desirable to construct a shell theory without these {\it ad hoc} assumptions/simplifications, which is consistent with the 3D formulation (field equations and top/bottom traction condition and incompressibility constraint)  to a proper asymptotic order in a pointwise manner.

We also mention that by the $\Gamma$-convergence method, Li and Chermisi \cite{li2013karman} rigorously derived the von K\'{a}rm\'{a}n shell theory for incompressible hyperelastic materials. However, this kind of approach depends on some {a priori} scaling assumptions, which cannot  yield a shell theory with both stretching and bending effects.

In a recent paper of Dai and Song \cite{dai2014consistent}, a dimension-reduction method was proposed to construct a consistent plate theory with both stretching and bending effects via series expansions with only smoothness assumption (without any {\it ad hoc} kinematic or other assumptions). The idea is to directly work with the 3D field equations and traction conditions on the top and bottom surfaces, and then to establish some recurrence relations for the expansion coefficients. Then, the approach has been used to derive a dynamic plate theory \cite{song2016consistent}, a static shell theory \cite{song2016consistent2}, a static plate theory for incompressible materials \cite{wang2016consistent} and a static shell theory for incompressible materials \cite{li2019consistent}.

In this paper, we follow Dai and Song's approach to first derive one form of dynamic shell theory for incompressible hyperelastic materials that involves three shell constitutive relations and six boundary conditions at each edge point. The completely new part is on the further refinement by elaborate calculations (cf. the procedure for a plate in \cite{wang2019uniformly}), which reduces the number of shell constitutive relations to two and singles out the bending term. It turns out that the refined shell equations alone can reveal a few new insights already. For the force boundary, in practice one only knows four conditions: the bending moment along the edge tangent direction and the three components of the cross-thickness resultant. To propose proper boundary conditions, we incorporate the weak form of the refined shell equations into the variation of the 3D Lagrange functional $\delta L$. By some elaborate calculations, which provide guidance on choosing the variation of the displacement vector in the 3D edge term in $\delta L$ when specializing to a  2D shell theory,  suitable shell boundary conditions and the 2D shell virtual work principle are obtained. A benchmark problem of an artery segment subjected to extension and internal pressure is considered. Finally, as an application of the refined shell theory, the plane-strain vibrations of a pressurized  artery are studied, and the results reveal the influences of the axial pre-stretch, pressure, and fibre angle on the vibration frequencies.

{\bf Notation.} Throughout this paper, we use boldface letters to denote vectors and second-order tensors; we use curly letters to denote higher-order tensors. The summation convention for repeated indices is adopted. In a summation, Greek letters  $\alpha,\beta,\gamma,\dots$ run from $1$ to $2$, whereas Latin letters $i,j,k,\dots$ run from $1$ to $3$. A comma preceding indices means differentiation and a dot over variables indicates time derivative. The time argument in variables is usually omitted for brevity. 

Let $\mathbb{R}^3$ be the three-dimensional Euclidean space with standard basis $(\bm{e}_1,\bm{e}_2,\bm{e}_3)$. The symbol $\bm{I}:=\bm{e}_i\otimes\bm{e}_i$ is reserved for the identity tensor of $\mathbb{R}^3$. The notation $\wedge$ means cross product. For a scalar-valued function of a tensor $W(\bm{F})$, the derivative of the $W$ with respect to $\bm{F}$ is defined to be $\frac{\partial W}{\partial\bm{F}}:=\frac{\partial W}{\partial F_{ji}}\bm{e}_i\otimes\bm{e}_j$; higher-order derivatives are defined in a similar way. The divergence of a tensor $\bm{S}$ is defined by $
\Div(\bm{S}):=\frac{\partial S_{ij}}{\partial x_i}\bm{e}_j$. The tensor contractions are defined by
\begin{align}
&\bm{A}[\bm{B}]=\tr(\bm{A}\bm{B}):=A_{ji}B_{ij},\quad \mathcal{A}^1[\bm{A}]:=\mathcal{A}_{ij\ell k}A_{k\ell}\bm{e}_i\otimes\bm{e}_j,\quad \bm{A}[\bm{a},\bm{b}]:=\bm{A}\bm{a}\cdot\bm{b}=A_{ij}a_jb_i.
\end{align}

\section{Kinematics and the 3D formulation}\label{sec:kinematics}
We consider a thin shell of constant thickness $2h$ composed of an incompressible hyperelastic material which occupies a region $\Omega\times [0,2h]$ in the reference configuration. The thickness $2h$ of the shell is assumed to be small compared with the length scale of the bottom surface $\Omega$ and its ratio against the radius of curvature is less than $1$. The position of a material point is denoted by $\bm{X}$ in the reference configuration and by $\bm{x}$ in the current configuration. The geometric description of a shell has been given in \cite{ciarlet2005introduction} and \cite{steigmann2012extension}, and here we give a brief summary.

The bottom surface $\Omega$ of the shell is parameterized by two curvilinear coordinates $\theta^\alpha$, $\alpha=1,2$. The position of a point on $\Omega$ is written as $\bm{r}=\bm{r}(\theta^\alpha)$. Then the tangent vectors along the coordinate lines are given by $\bm{g}_\alpha=\partial \bm{r}/\partial \theta^\alpha$, which form a covariant basis of the tangent plane of the bottom surface. Their contravariant counterparts $\bm{g}^\alpha$, which satisfy the relations $\bm{g}^{\alpha}\cdot\bm{g}_\beta=\delta^\alpha_\beta$,	
form a contravariant basis of the same plane. The unit normal vector $\bm{n}$ to the bottom surface is defined via $\bm{n}={\bm{g}_1\wedge \bm{g}_2}/{|\bm{g}_1\wedge \bm{g}_2|}$, so that  by setting $\bm{g}^3=\bm{g}_3=\bm{n}$, $\{\bm{g}_i\}$ and $\{\bm{g}^i\}$, $i=1,2,3$ form two sets of right-handed bases.

In the reference configuration, the position of a material point is decomposed into
\begin{equation}\label{eq:X}
\bm{X} = \bm{r}(\theta^\alpha) + Z \bm{n}(\theta^\alpha),\ \ 0\leq Z\leq 2h,
\end{equation}
where $Z$ is the coordinate of the point along the normal direction $\bm{n}$. The change of the unit normal vector  is captured by the curvature map, which is defined as the negative of the tangent map of the Gauss map $\bm{n}:\Omega\to S^2$ \cite{ciarlet2005introduction}, where $S^2$ denotes the two-dimensional unit sphere; thus we have $\bm{k}=-\partial \bm{n}/\partial \bm{r}=-\bm{n}_{,\alpha}\otimes\bm{g}^\alpha$.
We point out that the curvature tensor $\bm{k}$ is symmetric in the sense that $\bm{k}=\bm{k}^T$. 
Associated to $\bm{k}$, the mean curvature and the Gaussian curvature are respectively defined  by $H =\frac{1}{2}\tr(\bm{k})$ and $ K=\det(\bm{k})$.

The covariant basis vectors at a point in the shell $\Omega\times [0,2h]$ are given by
\begin{equation}\label{eq:galpha}
\widehat{\bm{g}}_\alpha =\frac{\partial \bm{X}}{\partial\theta^\alpha}=\frac{\partial\bm{r}}{\partial \theta^\alpha}+Z \frac{\partial \bm{n}}{\partial \bm{r}}\frac{\partial\bm{r}}{\partial \theta^\alpha}=(\bm{1}-Z\bm{k})\bm{g}_\alpha,
\end{equation} 
where $\bm{1}:=\bm{I}-\bm{n}\otimes\bm{n}=\bm{g}^\alpha\otimes\bm{g}_\alpha$ denotes the projection onto the tangent plane of $\Omega$; it is also the identity map of the same plane. Setting $\bm{\mu}=\bm{1}-Z\bm{k}$, we see from \eqref{eq:galpha} that 
$\widehat{\bm{g}}_\alpha =\bm{\mu}\bm{\bm{g}}_\alpha$ and thus $\widehat{\bm{g}}^\alpha =\bm{\mu}^{-T}\bm{\bm{g}}^\alpha$.
Note that the previous geometric assumption which asserts $|2hk_\alpha^\beta|< 1$ implies that the inverse   $\bm{\mu}^{-1}$ is well-defined. By the  change of variables formula, the volume element of the shell is computed by
\begin{equation}
dV = (\widehat{\bm{g}_1}\wedge \widehat{\bm{g}}_2)\cdot \bm{n}\, d\theta^1 d\theta^2 dZ =\det(\bm{\mu})(\bm{g}_1\wedge \bm{g}_2)\cdot \bm{n}\, d\theta^1 d\theta^2 dZ=\mu(Z)\,dAdZ,
\end{equation}
where $\mu(Z)=\det(\bm{\mu})= 1-2HZ+KZ^2$ and $dA=|\bm{g}_1\wedge \bm{g}_2|\,d\theta^1d\theta^2$ is the area element on the bottom surface.

On the boundary $\partial\Omega$, let $s$ be the arc length variable, and let $\bm{\tau}$ and $\bm{\nu}$ be respectively the unit tangent vector and the unit outward normal vector such that $(\bm{\tau},\bm{n},\bm{\nu})$ forms a right-handed triple (i.e., $\bm{\nu}=\bm{\tau}\wedge\bm{n}$). Then let $\bm{N}$, $\bm{T}$ and $da$ be respectively the unit outward normal vector, unit tangent vector and the area element of the lateral surface such that $(\bm{T},\bm{n},\bm{N})$ forms a right-handed triple. A similar argument as in \eqref{eq:galpha} yields $\bm{T}=(\bm{1}-Z\bm{k})\bm{\tau}/\sqrt{g_\tau}$, where $\sqrt{g_\tau}$ denotes the magnitude of vector $(1-Z\bm{k})\bm{\tau}$ and is given by $\sqrt{g_\tau}=\sqrt{1-2 Z\bm{k}\bm{\tau}\cdot\bm{\tau}+Z^2\bm{k}\bm{\tau}\cdot \bm{k}\bm{\tau}}$. Using the  change of variables formula again,  we have $\bm{N}\,da=\bm{\mu}\bm{\tau}\,ds\wedge \bm{n}\,dZ=(\bm{1}-Z\bm{k})\bm{\tau}\wedge\bm{n}\,dsdZ$. Then from the equality $(\bm{k}\bm{\tau})\wedge\bm{n} =\tr(\bm{k})(\bm{\tau}\wedge\bm{n})-\bm{k}(\bm{\tau}\wedge\bm{n})$,
we deduce that
\begin{align}\label{eq:ledge}
\bm{N}\,da = (\bm{1}+Z(\bm{k}-2H\bm{1}))\bm{\nu} \,dsdZ.
\end{align}
Since $(1-Z\bm{k})\bm{\tau}=\sqrt{g_\tau}\bm{T}$ and $(\bm{T},\bm{n},\bm{N})$ forms a right-handed triple of unit vectors, we have $da=\sqrt{g_\tau}dsdZ$ and $\sqrt{g_\tau}\bm{N}=(\bm{1}+Z(\bm{k}-2H\bm{1}))\bm{\nu}$ from the above equations.

The deformation gradient is then calculated by
\begin{equation}\label{eq:gradient}
\bm{F} = \frac{\partial\bm{x}}{\partial\bm{X}} = \frac{\partial\bm{x}}{\partial \theta^\alpha}\otimes\widehat{\bm{g}}^\alpha +\frac{\partial \bm{x}}{\partial Z}\otimes\bm{n} =(\nabla\bm{x})\bm{\mu}^{-1}+\frac{\partial \bm{x}}{\partial Z}\otimes\bm{n},
\end{equation}
where $\nabla:= \frac{\partial}{\partial\theta^\alpha} \bm{g}^\alpha$ denotes the 2D gradient operator on the base surface $\Omega$. We remark that for the 2D gradient operator, one has the following Stokes' theorem
\begin{align}
&\int_{\Omega}\nabla \cdot (\bm{1}\bm{a})\,dA =\int_{\partial\Omega} \bm{a}\cdot\bm{\nu}\,ds,\quad \int_{\Omega}\nabla \cdot (\bm{1}\bm{S})\,dA =\int_{\partial\Omega} \bm{S}^T \bm{\nu}\,ds \label{eq:stoke}
\end{align}
for a vector field $\bm{a}$ and a tensor field $\bm{S}$, respectively.

For an incompressible material, one has the following incompressibility constraint
\begin{equation}\label{eq:constraint}
R(\bm{F})=\det(\bm{F}) -1=0.
\end{equation}
Assume further that the material is hyperelastic with a strain energy function $W(\bm{F})$. Then
the associated elastic moduli are defined by $\mathcal{A}^{i}(\bm{F})=\frac{\partial^{i+1}W}{\partial \bm{F}^{i+1}},\ i=1,2,\dots$. The strain energy function is assumed to satisfy the strong-ellipticity condition:
$(\mathcal{A}^1(\bm{F})[\bm{a}\otimes\bm{b}])[\bm{a}\otimes\bm{b}]>0$ for $\bm{a}\otimes \bm{b}\neq 0$.

Suppose that $\bm{q}^+$ and $\bm{q}^-$ are the external loads applied on the top and the bottom surfaces of the shell respectively. The boundary $\partial \Omega$ of the bottom surface $\Omega$ is divided into two parts: the position boundary $\partial\Omega_0$ subjected to the prescribed position $\bm{b}$ and the traction boundary $\Omega_q$ subjected to the applied traction $\bm{q}$. Then the kinetic energy $K$, the strain energy $S$, and the load potential $V$ of the shell are respectively given by
\begin{align}
K&=\int_\Omega\int_0^{2h}\frac{1}{2}\rho\dot{\bm{x}}\cdot \dot{\bm{x}}\mu(Z)\,dZdA,\quad  S=\int_\Omega\int_0^{2h} W(\bm{F})\mu(Z)\,dZdA,\\
\begin{split}
V&=-\int_{\Omega}(\bm{q}^-(\bm{r})\cdot \bm{x}(\bm{r},0)+\bm{q}^+(\bm{r})\cdot \bm{x}(\bm{r},2h)\mu(2h))\,dA\\ &\quad-\int_{\Omega}\int_0^{2h}\bm{q}_{b}\cdot \bm{x}\mu(Z)\,dZdA-\int_{\partial\Omega_q}\int_0^{2h}\bm{q}(s,Z)\cdot\bm{x}(s,Z)\,da,
\end{split}
\end{align}
where $\rho$ is the mass density of the shell, $\bm{q}_{b}$ is the body force and $da$ is the area element on the lateral surface $\partial\Omega\times [0,2h]$.

By Hamilton's principle, the 3D  equations are obtained when the energy functional $E=K+S+V$ attains its minimum under the constraint condition \eqref{eq:constraint}. Therefore we are led to consider the  Lagrange functional
\begin{equation}
L(\bm{x}(\bm{X}),p(\bm{X}))=K+S+V -\int_{\Omega}\int_0^{2h} p(\bm{X})R(\bm{F})\mu(Z)\,dZdA,
\end{equation}
where $p(\bm{X})$ is the Lagrange multiplier.  To attain the minimum, it is necessary that the variation of $L$ with respect to $\bm{x}$ is zero, and a direct calculation shows
\begin{align}\label{eq:variation}
\begin{split}
\delta L&=\int_{\Omega}\int_0^{2h}(\rho\ddot{\bm{x}}-\Div(\bm{S})-\bm{q}_{b})\cdot \delta\bm{x}\mu(Z)\,dZdA -\int_{\Omega} (\bm{S}^T\bm{n}|_{Z=0}+\bm{q}^-)\cdot\delta\bm{x}(
\bm{r},0)\,dA\\
&\quad +\int_{\Omega} (\bm{S}^T\bm{n}|_{Z=2h}-\bm{q}^+)\cdot\delta\bm{x}(\bm{r},2h)\mu(2h)\, dA 
+\int_{\partial\Omega_q}\int_0^{2h} (\bm{S}^T\bm{N}-\bm{q})\cdot\delta\bm{x}(s,Z)\,da=0,
\end{split}
\end{align} 
where 
\begin{align}\label{eq:Sdefintion}
\bm{S}=\frac{\partial W}{\partial \bm{F}} -p\bm{F}^{-1}
\end{align}
is the nominal stress tensor of the incompressible hyperelastic material \cite{ogden1997non}, and we used ${\partial R}/{\partial\bm{F}}=\det(\bm{F})\bm{F}^{-1}$ and $\det(\bm{F})=1$. Since $\delta\bm{x}$ in \eqref{eq:variation} is arbitrary, we obtain the following 3D momentum equations together with boundary conditions: 
\begin{align}
&\Div (\bm{S})+\bm{q}_{b}=\rho\ddot{\bm{x}}\quad \text{in}\ \Omega\times [0,2h], \label{eq:kinematic}\\
&\bm{S}^T\bm{n}|_{Z=0}=-\bm{q}^-,\quad \bm{S}^T\bm{n}|_{Z=2h}=\bm{q}^+\quad \text{in}\ \Omega, \label{eq:botop}  \\
&\bm{S}^T\bm{N}=\bm{q}(s,Z)\quad \text{on}\ \partial\Omega_q\times [0,2h], \label{eq:lateral0}\\
&\bm{x}=\bm{b}(s,Z)\quad \text{on}\ \partial\Omega_0\times [0,2h]. \label{eq:lateralq}
\end{align}
The above equations together with the incompressibility constraint \eqref{eq:constraint} form the 3D dynamic equations for the shell structure, which contain an independent vector variable $\bm{x}$ and an independent scalar variable $p$.

\section{Refined 2D dynamic shell equations}\label{sec:refined 2D}

In this section, we shall first derive one form of consistent shell equations with three shell constitutive relations. Here the consistency means each term in \eqref{eq:variation} should be of a required asymptotic order, separately for the approximation. Then, a refinement is performed to reduce the number of shell constitutive relations from three to two. Also, the bending term is singled out. For the first part, the derivation is similar to that of the static case \cite{li2019consistent}, but to be self-contained, we present the main steps.

\subsection{Derivation of one form of 2D dynamic shell equations}

We assume sufficient smoothness for the quantities involved. Then $\bm{x}(\bm{X}), p(\bm{X}), \bm{F}(\bm{X})$ and $\bm{S}(\bm{X})$ have Taylor expansions about the bottom surface $Z=0$. From \eqref{eq:gradient} and the nonlinear relation \eqref{eq:Sdefintion}, the following relations among their expansion coefficients can be found:  
\begin{align}\label{eq:F}
&\bm{F}^{(0)}=\nabla\bm{x}^{(0)}+\bm{x}^{(1)}\otimes\bm{n},\quad  \bm{F}^{(1)}=\nabla\bm{x}^{(0)}\bm{k}+\nabla\bm{x}^{(1)}+\bm{x}^{(2)}\otimes\bm{n}, 
\end{align}
and 
\begin{align}\label{eq:S}
&\bm{S}^{(0)}  = \bm{A}^{0}-p^{(0)}\bm{F}^{(0)-1},\quad \bm{S}^{(1)}  = \mathcal{A}^{1}[\bm{F}^{(1)}]+p^{(0)}\bm{F}^{(0)-1}\bm{F}^{(1)}\bm{F}^{(0)-1}-p^{(1)}\bm{F}^{(0)-1},
\end{align}
where the superscript $^{(i)}$ denotes the $i$th derivative with respect to $Z$ at $Z=0$, and $\bm{A}^{0}=\partial W/\partial\bm{F}|_{\bm{F}=\bm{F}^{(0)}}$. 
From the above expressions, one easily checks that $\bm{S}^{(i)}$ is linear algebraic in $p^{(i)}$ and $\bm{x}^{(i+1)}$, $i=1$ (also true for $i=2$; for brevity the relations for $\bm{F}^{(2)}$ and $\bm{S}^{(2)}$ are omitted). It is due to this linearity that some recurrence relations can be established for the expansion coefficients upon further using the field equations in the subsequent derivations.

\begin{remark}
	The expressions for $\bm{S}^{(i)} (i=0,1,2)$ give three relations between the stress coefficients and the position vector coefficients. In the sequel, we abuse the terminology a little and call equations $\eqref{eq:S}_1$ and $\eqref{eq:S}_2$ and that for $\bm{S}^{(2)}$ shell constitutive relations. The reason is that the derived shell equations are represented in terms of $\bm{S}^{(i)}$ and through these relations the unknown in the shell equations is actually the position vector $\bm{x}^{(0)}$.
\end{remark}

Now, we shall proceed to do the dimension reduction process by using the 3D formulation. First, the bottom traction condition $\eqref{eq:botop}_1$ yields
\begin{equation}\label{eq:S0}
\bm{S}^{(0)T}\bm{n}=(\bm{A}^{0}-p^{(0)}\bm{F}^{(0)-1})^T\bm{n}=-\bm{q}^-.
\end{equation}
To ease notation, we introduce the vector $\bm{g}=\bm{g}(\bm{x}^{(0)})=\det(\bm{F}^{(0)})\bm{F}^{(0)-T}\bm{n}=\bm{x}^{(0)}_{,1}\wedge \bm{x}^{(0)}_{,2}/|\bm{g}_1\wedge\bm{g}_2|$ (see \cite{li2019consistent}). Then noting $\eqref{eq:F}_1$ and $\det(\bm{F}^{(0)})=1$, the above equation can be written as
\begin{equation}\label{eq:A0}
(\bm{A}^{0}(\nabla\bm{x}^{(0)}+\bm{x}^{(1)}\otimes\bm{n}))^T\bm{n} = -\bm{q}^-+p^{(0)}\bm{g},
\end{equation}

Next, substituting the Taylor expansion for $\bm{S}$ into the field equation \eqref{eq:kinematic} and equating the coefficients of $Z^i$ ($i=0,1,\dots$) on both sides, we have
\begin{align}
&\nabla\cdot \bm{S}^{(0)} +\bm S^{(1)T}\bm{n} + \bm{q}_{b}^{(0)}=\rho\ddot{\bm{x}}^{(0)},\label{eq:keq1}\\
& \nabla\cdot \bm{S}^{(1)} + \bm S^{(2)T}\bm{n}+ (\bm{k}\bm{g}^\alpha)\cdot \bm{S}^{(0)}_{,\alpha} + \bm{q}_{b}^{(1)}=\rho\ddot{\bm{x}}^{(1)},\label{eq:keq2}
\end{align}
where $\nabla\cdot\bm{S}:=\bm{g}^\alpha\cdot\bm{S}_{,\alpha}$ denotes the 2D divergence of the tensor $\bm{S}$. Then substituting the Taylor expansion for $\bm{F}$ into the constraint equation \eqref{eq:constraint} and equating the coefficients of $Z^i$ to be zero, we obtain
\begin{align}
&\bm{g}\cdot \bm{x}^{(1)}-1=0,\label{eq:coneq1}\\  
&\bm{g}\cdot\bm{x}^{(2)}+\tr(\bm{F}^{(0)-1}(\nabla\bm{x}^{(0)}\bm{k}+\nabla\bm{x}^{(1)}))=0,\label{eq:coneq2} 
\end{align}
where in \eqref{eq:coneq1} we have used the equality $\bm{F}^{(0)-1}\bm{x}^{(1)}=\bm{n}$ implied by $\eqref{eq:F}_1$. By the way, we point out that there is a typo in $(28)_1$ of \cite{li2019consistent}.

With the use of $\eqref{eq:S}_2$, equation \eqref{eq:keq1} can be simplified into
\begin{equation}\label{eq:kineq1}
\bm{B}\bm{x}^{(2)} +\bm{f}_2 -p^{(1)}\bm{g}=\rho\ddot{\bm{x}}^{(0)}
\end{equation}
by defining
\begin{align}
&\bm{B}\bm{a} = (\mathcal{A}^{1}[\bm{a}\otimes \bm{n}]+p^{(0)}\bm{F}^{(0)-1}(\bm{a}\otimes\bm{n})\bm{F}^{(0)-1})^T\bm{n} \iff {B}_{ij}=\mathcal{A}^{1}_{3i3j}+p^{(0)}F^{(0)-1}_{3i}F^{(0)-1}_{3j},\\
&\bm{f}_2 =  (\mathcal{A}^{1}[\nabla\bm{x}^{(0)}\bm{k}+\nabla\bm{x}^{(1)}]+p^{(0)}\bm{F}^{(0)-1}(\nabla\bm{x}^{(0)}\bm{k}+\nabla\bm{x}^{(1)})\bm{F}^{(0)-1})^T\bm{n}+\nabla\cdot\bm{S}^{(0)} +\bm{q}^{(0)}_{b}.
\end{align}
From \eqref{eq:coneq2} and \eqref{eq:kineq1}, we obtain 
\begin{align}
&p^{(1)}  = \frac{1}{\bm{g} \cdot \bm{B}^{-1}\bm{g}}( \bm{g}\cdot \bm{B}^{-1}(\bm{f}_2-\rho\ddot{\bm{x}}^{(0)}) - \tr(\bm{F}^{(0)-1}(\nabla\bm{x}^{(0)}\bm{k}+\nabla\bm{x}^{(1)}))),\label{eq:p1}\\
&\bm{x}^{(2)}  = \bm{B}^{-1}(p^{(1)}\bm{g}+\rho\ddot{\bm{x}}^{(0)}-\bm{f}_2).\label{eq:x2}
\end{align}
Note that the strong-ellipticity condition guarantees that $\bm{B}$ is positive definite and hence is invertible. The explicit expressions of $\bm{x}^{(3)}$ and  $p^{(2)}$ can be obtained similarly, whose expressions are omitted. The explicit expressions of $\bm{x}^{(4)}$ and $p^{(3)}$ are not needed since they are intermediate variables. 
The explicit expressions for $\bm{x}^{(1)}$ and $p^{(0)}$ are encoded in  \eqref{eq:A0} and \eqref{eq:coneq1}, which are nonlinear algebraic equations in general, so they can only be solved when the strain energy function is specified. Nevertheless, the strong-ellipticity condition together with the implicit function theorem ensures that $\bm{x}^{(1)}$ and $p^{(0)}$ can be uniquely solved in terms of $\bm{x}^{(0)}$ (cf. \cite{wang2016consistent}).

Finally, the top traction condition $\eqref{eq:botop}_2$ states
\begin{equation}\label{eq:final}
\bm{S}^{(0)T}\bm{n} + 2h \bm{S}^{(1)T}\bm{n} + 2h^2 \bm{S}^{(2)T}\bm{n} + \frac{4}{3}h^3 \bm{S}^{(3)T}\bm{n} +O(h^4\bm{S}^{(4)T}\bm{n})=\bm{q}^+.
\end{equation}
Subtracting $\eqref{eq:final}$ multiplied by $\mu(2h)=1-4Hh+4Kh^2$ from $\eqref{eq:S0}$  and then simplifying (see \cite{song2016consistent} for details), we arrive at one form of a 2D dynamic vector shell equation
\begin{equation}\label{eq:compact}
\nabla \cdot \widetilde{\bm{S}}+O(h^3\bm{S}^{(3)},h^3k\bm{S}^{(2)})=\rho\ddot{\widetilde{\bm x}} -\widetilde{\bm{q}}+O(h^3\ddot{\bm{x}}^{(3)},h^3k\ddot{\bm{x}}^{(i)},h^3\bm{q}_b^{(3)},h^3k\bm{q}_b^{(i)}),
\end{equation}
where $i=1,2$ and
\begin{align}
\begin{split}\label{eq:Stilde}
\widetilde{\bm{S}}&=(\bm{1}+h(\bm{k}-2H\bm{1}))\bm{S}^{(0)}+h(\bm{1}+\frac{4}{3}h(\bm{k}-2H\bm{1}))\bm{S}^{(1)}+\frac{2}{3}h^2\bm{1}\bm{S}^{(2)}\\
&=\frac{1}{2h}\int_{0}^{2h} (\bm{1}+Z(\bm{k}-2H\bm{1}))\bm{S}\,dZ+O(h^3\bm{S}^{(3)},h^3k\bm{S}^{(2)}),
\end{split}\\
\begin{split}\label{eq:xtilde}
\widetilde{\bm{x}}
&=(1-2hH+\frac{4}{3}h^2K)\bm{x}^{(0)} +h(1-\frac{8}{3}hH)\bm{x}^{(1)}+\frac{2}{3}h^2\bm{x}^{(2)}\\
&=\frac{1}{2h}\int_0^{2h} \bm{x}\mu(Z)\,dZ+O(h^3\bm{x}^{(3)},h^3k\bm{x}^{(i)}),
\end{split}\\
\widetilde{\bm{q}}&  =\frac{\mu(2h)\bm{q}^+ +\bm{q}^-}{2h}+\widetilde{\bm{q}}_b,
\end{align}
and $\widetilde{\bm{q}}_b$ is defined in the same way as $\widetilde{\bm{x}}$.
\begin{remark}
	The quantity $\widetilde{\bm{S}}$ is considered as the averaged stress, and $\widetilde{\bm{q}}$ the averaged shell body force due to surface traction and 3D body force. We point out that \eqref{eq:compact} can be also deduced by multiplying the field equation \eqref{eq:kinematic} by $\mu(Z)$ and then integrating it with respect to $Z$ from $0$ to $2h$ followed by applying the equality
	\begin{align}\label{eq:2D}
	\int_{0}^{2h} \Div({\bm{S}})\mu(Z)\,dZ=\nabla\cdot(\int_{0}^{2h} (\bm{1}+Z(\bm{k}-2H\bm{1}))\bm{S}\,dZ)+\bm{S}^T\bm{n}|_{Z=2h}\mu(2h)-\bm{S}^T\bm{n}|_{Z=0},
	\end{align}
	which  is a consequence of Stokes' theorem.
\end{remark}

Similar to \cite{song2016consistent2}, suitable edge boundary conditions can be imposed, and then it can be shown that each of the five terms in \eqref{eq:variation} is of $O(h^4)$, which satisfies the consistency criterion. The details are omitted. Also, it is clear from the derivation process that the bottom traction condition, the 3D field equations, the incompressibility condition and the top traction condition are all satisfied in a pointwise manner (with an error of $O(h^4)$, see \eqref{eq:final}), an important feature not enjoyed by shell theories based on {\it ad hoc} assumptions and/or cross-thickness integrations.

\subsection{Refined 2D dynamic shell equations}\label{subsec:refined}

Although the above-derived shell theory is consistent, there are still a few undesirable features as follows. 1. There are a little too many (three) shell constitutive relations (equations $\eqref{eq:S}_1$ and $\eqref{eq:S}_2$ and that for $\bm{S}^{(2)}$). In particular, the relation between $\bm{S}^{(2)}$ and $\bm{x}^{(0)}$ is very complicated and can cause some technical difficulties for implementation in concrete applications. 2. From the shell equations, one cannot tell clearly which term(s) represents the bending effect. 3. Although the associated weak form can be obtained from the shell equations, physically  it does not represent the shell virtual work principle. 4. The shell equations are three coupled fourth-order PDEs for $\bm{x}^{(0)}$, which require six boundary conditions at an edge point. If one knows the displacement and/or stress distributions, there is no difficulty imposing them. However, in many practical situations for the traction edge, one only knows four conditions: the cross-thickness force resultant and the bending moment (with direction along the edge tangent), and one does not know how to impose the other two boundary conditions.  For a plate theory, these issues were addressed in \cite{wang2019uniformly}. Here, with some modifications, those ideas from this previous work will be used for a shell theory. In this subsection, we shall  resolve the first two issues by performing some manipulations to eliminate $\bm{S}^{(2)}$ and to single out the bending term. As a price to pay, the relative errors for some problems may not be as good as before. We point out that one cannot simply drop $\frac{2}{3}h^2\bm{1}\bm{S}^{(2)}$ in \eqref{eq:Stilde}, as the bending effect is also dropped. So, one needs to do some elaborate calculations to extract the bending term first and then to drop the relative higher-order terms. The last two issues will be resolved in the next section.

First, we rewrite \eqref{eq:compact} into two parts:
\begin{align}
& \bm{1}\nabla\cdot \widetilde{\bm{S}}+O(h^3\bm{S}^{(3)},h^3k\bm{S}^{(2)})=\rho\ddot{\widetilde{\bm{x}}}_{t}-\widetilde{\bm{q}}_{t}+O(h^3\ddot{\bm{x}}^{(3)}_t,h^3k\ddot{\bm{x}}^{(i)}_t,h^3\bm{q}^{(3)}_{bt},h^3k\bm{q}^{(i)}_{bt}),\label{eq:alpha12}\\
&
(\nabla\cdot\widetilde{\bm{S}})\cdot\bm{n}+O(h^3\bm{S}^{(3)},h^3k\bm{S}^{(2)})=\rho\ddot{\widetilde{x}}_{3}-\widetilde{q}_{3}+O(h^3\ddot{x}^{(3)}_3,h^3k\ddot{x}^{(i)}_3,h^3q^{(3)}_{b3},h^3kq^{(i)}_{b3}),\label{eq:alpha3}
\end{align}
where $\bm{1}=\bm{I}-\bm{n}\otimes\bm{n}=\bm{g}^\alpha\otimes \bm{g}_\alpha$ and the subscript $t$ indicates the projection into the tangent plane; thus $\bm{a}_t:=\bm{1}\bm{a}=\bm{a}\bm{1}$ and $\bm{S}_{t}:=\bm{1}\bm{S}\bm{1}$ for a vector $\bm{a}$ and a tensor $\bm{S}$ respectively. Note that since $\widetilde{\bm{S}}$ satisfies the equality  $\bm{1}\widetilde{\bm{S}}=\widetilde{\bm{S}}$ (see \eqref{eq:Stilde}), we have 
\begin{equation}\label{eq:St}
\widetilde{\bm{S}}_t=\bm{1}\widetilde{\bm{S}}\bm{1}=\widetilde{\bm{S}}\bm{1}.
\end{equation}

Next, we want to extract terms related to in-plane stress $\widetilde{\bm{S}}_t$ from the in-plane equation \eqref{eq:alpha12} in order to gain some insights as well for later use for deriving the 2D shell virtual work principle. For this purpose, we need the following two equalities for a tensor field $\bm{S}$ and a vector field $\bm{a}$:
\begin{align}
&\bm{1} \nabla\cdot \bm{S}=\bm{1}\nabla\cdot(\bm{S}\bm{1})-k^\alpha_\beta S^{\beta 3}\bm{g}_\alpha,\label{eq:S1}\\
&(\nabla\cdot \bm{S})\cdot\bm{a}=\nabla\cdot(\bm{S}\bm{a})-\tr(\nabla\bm{a}\bm{S}).\label{eq:Sa}
\end{align}	
To prove \eqref{eq:S1}, it suffices to show that
\begin{align}
\bm{1}\nabla\cdot(\bm{S}-\bm{S}\bm{1})=-k^\alpha_\beta S^{\beta 3}\bm{g}_\alpha.
\end{align}
Since $\bm{1}=\bm{I}-\bm{n}\otimes\bm{n}$, we have $\bm{S}-\bm{S}\bm{1}=\bm{S}\bm{n}\otimes\bm{n}$. Further, we have
\begin{align}
\bm{1}\nabla\cdot (\bm{S}\bm{n}\otimes\bm{n})&=\bm{1} (\bm{g}^\beta \cdot(\bm{S}\bm{n}\otimes\bm{n})_{,\beta})=\bm{g}^\beta\cdot(\bm{S}\bm{n})_{,\beta}\bm{1}\bm{n}+(\bm{g}^\beta\cdot\bm{S}\bm{n})\bm{1}\bm{n}_{,\beta}\\
&=-(\bm{g}^\beta\cdot\bm{S}\bm{n})\bm{k}\bm{g}_\beta=-k^\alpha_\beta S^{\beta 3}\bm{g}_\alpha.
\end{align}
Thus \eqref{eq:S1} follows.  Equation \eqref{eq:Sa} can be proved by a direct calculation starting from $\nabla\cdot(\bm{S}\bm{a})$ by using the definition of the 2D divergence.

Using \eqref{eq:S1}, \eqref{eq:Sa} and \eqref{eq:St}, and noting that $\nabla\bm{n}=-\bm{k}$, \eqref{eq:alpha12} and \eqref{eq:alpha3} can be rewritten as
\begin{align}
& \bm{1}\nabla\cdot\widetilde{\bm{S}}_t-k^\alpha_\beta\widetilde{S}^{\beta 3}\bm{g}_\alpha+O(h^3\bm{S}^{(3)},h^3k\bm{S}^{(2)})=\rho\ddot{\widetilde{\bm{x}}}_{t}-\widetilde{\bm{q}}_{t}+O(h^3\ddot{\bm{x}}^{(3)}_t,h^3k\ddot{\bm{x}}^{(i)}_t,h^3\bm{q}^{(3)}_{bt},h^3k\bm{q}^{(i)}_{bt}),\label{eq:alpha12b}\\
&
\nabla\cdot(\widetilde{\bm{S}}\bm{n})+\tr(\bm{k}\widetilde{\bm{S}}_t)+O(h^3\bm{S}^{(3)},h^3k\bm{S}^{(2)})=\rho\widetilde{x}_{3}-\widetilde{q}_{3}+O(h^3\ddot{x}^{(3)}_3,h^3k\ddot{x}^{(i)}_3,h^3q^{(3)}_{b3},h^3kq^{(i)}_{b3}).\label{eq:alpha3b}
\end{align}

Now, we shall manipulate the third equation \eqref{eq:alpha3b} further to single out the bending term. Adding \eqref{eq:final} multiplied by $\mu(2h)$ to $\eqref{eq:S0}$, we obtain
\begin{align}\label{eq:mmm}
\begin{split}
&(1-2hH+2h^2K)\bm{S}^{(0)T}\bm{n} + h(1-4hH) \bm{S}^{(1)T}\bm{n} + h^2\bm{S}^{(2)T}\bm{n}+O(h^3\bm{S}^{(3)T}\bm{n},h^3k\bm{S}^{(i)T}\bm{n})= \bm{m},
\end{split}
\end{align}
where  $i=1,2$ and $\bm{m}=(\mu(2h)\bm{q}^+-\bm{q}^-)/2$. To extract the bending term from \eqref{eq:alpha3b}, we subtract  the 2D divergence of \eqref{eq:mmm} multiplied by $\bm{1}$ from the left from  $\eqref{eq:alpha3b}$ (with the substitution of \eqref{eq:Stilde}). Note that the focus for this manipulation is on the $\bm{S}^{(2)}$ terms in these two equations. Then, upon further using \eqref{eq:S0} and \eqref{eq:keq2}, we obtain 
\begin{align}
\begin{split}\label{eq:mmmm}
&\quad\nabla\cdot((\bm{1}+h(\bm{k}-2H\bm{1}))\bm{S}^{(0)}\bm{n}-((1-2hH+2h^2K)\bm{1}+h \bm{k}) \bm{S}^{(0)T}\bm{n})\\
&\quad+h\nabla\cdot((\bm{1}+\frac{4}{3}h(\bm{k}-2H\bm{1}))\bm{S}^{(1)}\bm{n}-(1-4hH)\bm{1}\bm{S}^{(1)T}\bm{n})+\frac{2}{3}h^2\nabla\cdot (\bm{1}\bm{S}^{(2)}\bm{n}-\bm{1}\bm{S}^{(2)T}\bm{n})\\
&\quad + \tr(\bm{k}\widetilde{\bm{S}}_t)+\frac{1}{3}h^2\nabla\cdot(\bm{1}((\bm{k}\bm{g}^\alpha)\cdot\bm{S}^{(0)}_{,\alpha}))+\frac{1}{3}h^2\nabla\cdot(\bm{1}\nabla\cdot\bm{S}^{(1)})+O(h^3\bm{S}^{(3)},h^3k\bm{S}^{(i)}) \\
&=\rho \ddot{\widetilde{x}}_{3}-\widetilde{q}_3+\frac{1}{3}h^2\nabla\cdot(\rho\ddot{\bm{x}}^{(1)}_t-\bm{q}_{bt}^{(1)})-\nabla\cdot \bm{m}_{t}+h\nabla\cdot(\bm{k}\bm{q}^-_t)+O(h^3\ddot{x}^{(3)}_3,h^3k\ddot{x}^{(i)}_3,h^3q^{(3)}_{b3},h^3kq^{(i)}_{b3}).
\end{split}
\end{align}

We also want to extract the in-plane stress parts of the last term $\frac{1}{3}h^2\nabla\cdot(\bm{1}\nabla\cdot\bm{S})$ on the left-hand side. Observe that we have  the decomposition
\begin{align}
\begin{split}
\bm{S}^{(1)} &=\bm{I}\bm{S}^{(1)}\bm{I}=(\bm{1}+\bm{n}\otimes\bm{n})\bm{S}^{(1)}(\bm{1}+\bm{n}\otimes\bm{n})=\bm{S}^{(1)}_t+\bm{n}\otimes\bm{1}\bm{S}^{(1)T}\bm{n}+\bm{S}^{(1)}\bm{n}\otimes\bm{n}.
\end{split}
\end{align} 
Further,  routine calculations show that
\begin{align}
&\nabla\cdot(\bm{1}\nabla\cdot(\bm{n}\otimes\bm{1}\bm{S}^{(1)T}\bm{n}))=-\nabla\cdot(2H\bm{1}\bm{S}^{(1)T}\bm{n})),\\
&\nabla\cdot(\bm{1}\nabla\cdot (\bm{S}^{(1)}\bm{n}\otimes\bm{n}))=-\nabla\cdot ((\bm{g}^\alpha\cdot \bm{S}^{(1)}\bm{n})\bm{k}\bm{g}_\alpha).
\end{align}
Upon using the above three equations, \eqref{eq:mmmm} can be recast as
\begin{align}
\begin{split}\label{eq:mmmma}
&\quad\nabla\cdot((\bm{1}+h(\bm{k}-2H\bm{1}))\bm{S}^{(0)}\bm{n}-((1-2hH+2h^2K)\bm{1}+h \bm{k}) \bm{S}^{(0)T}\bm{n})\\
&\quad+h\nabla\cdot((\bm{1}+\frac{4}{3}h(\bm{k}-2H\bm{1}))\bm{S}^{(1)}\bm{n}-(1-4hH)\bm{1}\bm{S}^{(1)T}\bm{n})+\frac{2}{3}h^2\nabla\cdot (\bm{1}\bm{S}^{(2)}\bm{n}-\bm{1}\bm{S}^{(2)T}\bm{n})\\
&\quad + \tr(\bm{k}\widetilde{\bm{S}}_t)+\frac{1}{3}h^2\nabla\cdot(\bm{1}\nabla\cdot\bm{S}^{(1)}_t)
-\frac{1}{3}h^2 \nabla\cdot(2H\bm{1}\bm{S}^{(1)T}\bm{n})-\frac{1}{3}h^2\nabla\cdot((\bm{g}^\alpha\cdot\bm{S}^{(1)}\bm{n})\bm{k}\bm{g}_\alpha)\\
&\quad +\frac{1}{3}h^2\nabla\cdot(\bm{1}((\bm{k}\bm{g}^\alpha)\cdot\bm{S}^{(0)}_{,\alpha}))+O(h^3\bm{S}^{(3)},h^3k\bm{S}^{(i)})\\
&=\rho \ddot{\widetilde{x}}_{3}-\widetilde{q}_3+\frac{1}{3}h^2\nabla\cdot(\rho\ddot{\bm{x}}^{(1)}_t-\bm{q}_{bt}^{(1)})-\nabla\cdot \bm{m}_{t}+h\nabla\cdot(\bm{k}\bm{q}^-_t)+O(h^3\ddot{x}^{(3)}_3,h^3k\ddot{x}^{(i)}_3,h^3q^{(3)}_{b3},h^3kq^{(i)}_{b3}).
\end{split}
\end{align}

To eliminate $\bm{S}^{(2)}$ terms in a consistent manner, we shall drop any term which is relatively $O(h^2)$ or $O(h)$ smaller than another term (so that the shell theory yields results with a relative $O(h^2)$ or $O(h)$ error). It is justified, as shown in the following  simple example: for $A+B+C=0$, if $C=O(h^2 B)$ or $C=O(hB)$, the dropping of $C$ causes at most a  relative error of $O(h^2)$ or $O(h)$, no matter $A> O(B)$ or $A\le O(B)$. Any terms which cannot satisfy the above requirement will be kept.

We make the following observations. 1. In \eqref{eq:alpha12b},  $\frac{2}{3}h^2\bm{1}\bm{S}^{(2)}$ in $\widetilde{\bm{S}}$ (cf. \eqref{eq:Stilde}) is dropped, as it is $O(h^2)$ smaller than $\bm{1}\bm{S}^{(0)}$ or $O(h)$ smaller than $h\bm{1}\bm{S}^{(1)}$ if $\bm{S}^{(0)}=0$ (e.g., the bottom surface undergoes an inextensible rotation, for which $\bm{F}^{(0)}=\bm{R}$ and thus $\bm{S}^{(0)}=0$, where $\bm{R}$ is a rotation tensor). As it is possible that $\bm{S}^{(1)}$ terms become the leading ones, they should be kept. 2. The last three terms on the left-hand side of \eqref{eq:mmmma}, $h^2\nabla\cdot(2K\bm{1}\bm{S}^{(0)T}\bm{n})$, $\frac{4}{3}h^2\nabla\cdot((\bm{k}-2H\bm{1})\bm{S}^{(1)}\bm{n})$ and $h^2\nabla\cdot(4H\bm{1}\bm{S}^{(1)T}\bm{n})$ are dropped as they are $O(h^2)$ smaller than $\tr(\bm{k}\widetilde{\bm{S}}_t)$ or either  $O(h)$  smaller than $\tr(\bm{k}\widetilde{\bm{S}}_t)$ or zero if $\bm{S}^{(0)}=0$. 3. The third term on the left-hand side of \eqref{eq:mmmma}  is dropped as it is $O(h^2)$ smaller than $\nabla\cdot(\bm{1}\bm{S}^{(0)}\bm{n}- \bm{1}\bm{S}^{(0)T}\bm{n})$ or $O(h)$ smaller than $h\nabla\cdot  (\bm{1}\bm{S}^{(1)}\bm{n}- \bm{1}\bm{S}^{(1)T}\bm{n})$ if $\bm{S}^{(0)}=0$.  4. On the right-hand sides, $\frac{1}{3}h^2\bm{x}^{(2)}$ in $\widetilde{\bm{x}}$ (cf. \eqref{eq:xtilde}) is dropped , as it is $O(h^2)$ smaller than $\bm{x}^{(0)}$, and a similar treatment is  made to $\widetilde{\bm{q}}_b$.  From these observations, we have the refined 2D  dynamic shell equations as follows:
\begin{align}
&\bm{1}\nabla\cdot\overline{\bm{S}}_t-k^\alpha_\beta \overline{S}^{\beta 3}\bm{g}_\alpha=\rho \ddot{\overline{\bm{x}}}_t -\overline{\bm{q}}_t,\label{eq:ffinal12}\\
\begin{split}
&\nabla \cdot(\overline{\bm{S}_{\star}}\bm{n}-\overline{\bm{S}^T_{\star}}\bm{n})+\tr(\bm{k}\overline{\bm{S}}_t)+\frac{1}{3}h^2\nabla\cdot(\bm{1}\nabla\cdot\bm{S}^{(1)}_t)\\
=&\rho \ddot{\overline{x}}_{3} -\overline{q}_{3}+\frac{1}{3}h^2\nabla\cdot(\rho\ddot{\bm{x}}^{(1)}_t-\bm{q}_{bt}^{(1)})-\nabla\cdot\bm{m}_{t}+h\nabla\cdot(\bm{k}\bm{q}^-_t),\label{eq:ffinal3}
\end{split}
\end{align}
where 
\begin{align}
&\overline{\bm{S}}=(\bm{1}+h(\bm{k}-2H\bm{1}))\bm{S}^{(0)}+h(\bm{1}+\frac{4}{3}h(\bm{k}-2H\bm{1}))\bm{S}^{(1)},\\
&\overline{\bm{S}_{\star}}=(\bm{1}+h(\bm{k}-2H\bm{1}))\bm{S}^{(0)}+h\bm{1}\bm{S}^{(1)},\\
&\overline{\bm{S}^T_{\star}}=(\bm{1}+h(\bm{k}-2H\bm{1}))\bm{S}^{(0)T}+h\bm{1}\bm{S}^{(1)T},\\
&\overline{\bm{x}}=(1-2hH+\frac{4}{3}h^2K)\bm{x}^{(0)} +h(1-\frac{8}{3}hH)\bm{x}^{(1)},\\
&\overline{\bm{q}}=\frac{\mu(2h)\bm{q}^{+}+\bm{q}^-}{2h}+\overline{\bm{q}}_b,
\end{align}
and  $\overline{\bm{q}}_b$ is defined in the same way as $\overline{\bm{x}}$.

From the above shell equations, one can observe some important insights. 1. For a plate (or a shell with $|k^\alpha_\beta|\leq O(h^2)$) in linear elasticity, the bending term $\frac{1}{3}h^2\nabla\cdot(\bm{1}\nabla\cdot\bm{S}^{(1)}_t)$ becomes the leading term, so it should be kept although it looks like an $O(h^2)$ term. 2. For the in-plane equation \eqref{eq:ffinal12}, the in-plane forces and inertia effects are resisted by two sources: the in-plane stress part (the first term on the left-hand side) and the out-plane shear stresses due to the curvature effect (the second term). 3. For the out-plane equation \eqref{eq:ffinal3}, the out-plane forces and inertia effects are resisted by three sources: (i) the out-plane shear stresses (the first term on the left-hand side) due to geometric and/or material nonlinearity; (ii) the in-plane stresses due to the curvature effect (the second term); (iii) bending effect due to the in-plane stresses (the last term). 4. Although the out-plane normal stress does not appear explicitly in these shell equations, it plays a role in expressing $\bm{x}^{(1)}$ and $p^{(0)}$ in terms of $\bm{x}^{(0)}$ (see \eqref{eq:S0} and \eqref{eq:coneq1}), so it should not be ignored (as in some {\it ad hoc} theories, which assume the out-plane component of the displacement is independent of $Z$). 5. Only two shell constitutive relations are needed, which are provided by $\eqref{eq:S}_1$ and $\eqref{eq:S}_2$. 6. These shell equations provide  results with at most a relative $O(h)$ error, although in some cases the error can be $O(h^2)$. Note that higher-order Taylor expansions do not necessarily lead to higher-order correct plate/shell equations.

After substitutions of all recurrence relations, the above shell equations become a system of differential equations involving $\bm{x}^{(0)}$ only. Once it is solved,  $\bm{x}^{(0)}$ (with a relative error equal to or smaller than $O(h)$) is obtained and the position vector $\bm{x}$ can then be recovered.

\section{Boundary conditions and shell virtual work principle }\label{sec:shell virtual work principle}

Now we shall resolve the last two issues mentioned in the beginning of the previous subsection. Actually, boundary conditions for a derived shell theory can cause considerable difficulty (see Steigmann \cite{steigmann2013koiter}).  Here, we shall use both the variation of the 3D Lagrange functional and the weak form of the shell equations to get the appropriate boundary conditions and the 2D shell virtual work principle.

For the shell equations, the bottom traction condition $\eqref{eq:botop}_1$, and the vanishing coefficients of the field equation \eqref{eq:kinematic} and the incompressibility constraint \eqref{eq:constraint} are used to find the recurrence relations.  As a result, \eqref{eq:kinematic} (up  to required order) and $\eqref{eq:botop}_1$ can be treated as identities. To obtain the 2D shell virtual work principle from the vanishing of the variation of 3D Lagrange functional \eqref{eq:variation}, we need to specialize it to the 2D case (by using the Taylor expansions for the quantities involved as in deriving the shell equations).  The first two terms in \eqref{eq:variation} can be set to be identically zero because of the above-mentioned two identities. Then, in order to remove $\delta\bm{x}(\bm{r},2h)$ (we still use $\bm{x}(\bm{r},2h)$ for the writing purpose but it means the Taylor expansion of the position vector at $Z=2h$) in the third integral and to introduce $\delta\bm{x}(\bm{r},h)$ to the variation (needed for the 2D shell virtual work principle), we add to $\delta L$ three identically zero terms (the first three terms below) to obtain
\begin{align}
\begin{split}
\delta L=&2h\int_{\Omega} \bm{A}_t\cdot (\delta\bm{x}_{t}(\bm{r},2h)-\delta\bm{x}_t(\bm{r},h))\,dA+2h\int_{\Omega} A_3\cdot (\delta x_3(\bm{r},2h)-\delta x_3(\bm{r},h))\,dA\\
& +2h \int_{\Omega} (\nabla\cdot (\bm{1}\bm{C}))\cdot \delta\bm{x}(\bm{r},2h)\,dA+\int_{\Omega}(\bm{S}^{T}\bm{n}|_{Z=2h}-\bm{q}^+)\cdot\delta\bm{x}(\bm{r},2h)\mu(2h)\,dA\\
&+\int_{\partial\Omega_q}\int_{0}^{2h}(\bm{S}^T\bm{N}-\bm{q})\cdot\delta\bm{x}(s,Z)\,da=0,
\end{split}
\end{align}
where $\bm{A}_t=0$, $A_3=0$ and $\bm{C}=0$ correspond to equations \eqref{eq:ffinal12}, \eqref{eq:ffinal3} and \eqref{eq:mmm} respectively. Also, we remark that the last edge term is still of the 3D one and we delay  to specialize it to the 2D shell theory later. A direct calculation shows that the $\delta\bm{x}(\bm{r},2h)$ terms cancel each other (upon dropping relatively higher-order terms as in Section 3\ref{subsec:refined}), and thus we have
\begin{align}\label{eq:deltaL}
\delta L=&-2h\int_{\Omega} \bm{A}_t\cdot \delta\bm{u}_{mt}\,dA-2h\int_{\Omega} A_3\cdot \delta u_{m3}\,dA+\int_{\partial\Omega_q}\int_{0}^{2h}(\bm{S}^T\bm{N}-\bm{q})\cdot\delta\bm{u}(s,Z)\ da=0,
\end{align}
where we have used the virtual displacement $\delta \bm{u}$ to replace the virtual position vector and the subscript $m$ denotes the middle surface $Z=h$. Actually, the first two terms are just the weak form for the shell equations \eqref{eq:ffinal12} and \eqref{eq:ffinal3}.  We remark that when the boundary conditions are involved, one can only expect to obtain the leading-order results in general; thus in the sequel, any term, which is relatively smaller than another term, will be dropped.

To get the 2D shell virtual work principle, we shall further add two identities to the above equation, which are associated with the virtual work due to the moment, which is given by
\begin{align}
\bm{M}=\int_{\partial \Omega}\int_{0}^{2h} ((\bm{x}-\bm{x}(\bm{r},h))\times \bm{S}^T\bm{N})\sqrt{g_{\tau}}
\,  dZ ds.
\end{align}
Then, the twist moment (along $\bm{N}_m$ direction) and the bending moment (along $\bm{T}_m$ direction) per unit arc length of $\partial\Omega$ are given by respectively
\begin{align}
\begin{split}
T&=\int_{0}^{2h} ((\bm{x}-\bm{x}(\bm{r},h))\times \bm{S}^T\bm{N})\cdot \bm{N}_m\sqrt{g_\tau}\,dZ\\
&=\frac{2}{3}h^3\bm{S}^{(1)T}[\bm{\nu},\bm{\nu}\times\bm{x}^{(1)}]+\frac{1}{3}h^3\bm{S}^{(0)T}[\bm{\nu},\bm{\nu}\times\bm{x}^{(2)}]+O(h^4,h^3k),\label{eq:T}
\end{split}
\end{align}
\begin{align}
\begin{split}
M&=\int_{0}^{2h} ((\bm{x}-\bm{x}(\bm{r},h))\times \bm{S}^T\bm{N})\cdot \bm{T}_m\sqrt{g_\tau}\,dZ\\
&=\frac{2}{3}h^3\bm{S}^{(1)T}[\bm{\nu},\bm{\tau}\times\bm{x}^{(1)}]+\frac{1}{3}h^3\bm{S}^{(0)T}[\bm{\nu},\bm{\tau}\times\bm{x}^{(2)}]+O(h^4,h^3k).\label{eq:M}
\end{split}
\end{align}
It was shown in \cite{krauthammer2001thin} (Section 2.5; the authors attributed the argument to Kirchhoff) that the derivative of the twisting moment with respect to the arc length ${T}_{,s}$ is equivalent to a distributed shear force (along the downward thickness direction). Thus, this twist moment generates a virtual work per unit arc length: $-T_{,s}\delta u_{m3}$ (the smoothness of $\partial \Omega$ is assumed). On the other hand, the bending moment generates a virtual work per unit arc length: $-M \delta \alpha_m$, where $\alpha_m$ is the rotation angle at the edge of the middle surface. It is defined as the change of the angle between the vector $\bm{N}_m$ and the projected vector onto the $\bm{n}\bm{N}_m$-plane of the tangent vector at an edge point  of the intersection curve of the middle surface and the $\bm{n}\bm{N}_m$-plane during the deformation, which is given by (after some calculations)
\begin{align}\label{eq:rotation angle}
\alpha_m=\arctan(\frac{\nabla_m\bm{x}(\bm{r},h)[\bm{N}_m]\cdot\bm{n}}{\nabla_m\bm{x}(\bm{r},h)[\bm{N}_m]\cdot\bm{N}_m})\pm p \pi=\arctan(\frac{u_{m3,\nu}}{1+\bm{1}\nabla \bm{u}_{mt}[\bm{\nu},\bm{\nu}]}+O(k,hk))\pm p\pi,
\end{align}
where $\nabla_m=\frac{\partial }{\partial\theta^\alpha}\widehat{\bm{g}}_\alpha|_{Z=h}$ (see \eqref{eq:galpha} for the definition of $\widehat{\bm{g}}_\alpha$) is the gradient operator on the middle surface and $p$ is a natural number. 

Now, we add the two identities $-T_{,s} \delta u_{m3}+T_{,s} \delta u_{m3}=0$ and $-M \delta \alpha_m+M \delta \alpha_m=0$ to  equation \eqref{eq:deltaL} to obtain

\begin{align}\label{eq:dL}
\begin{split}
\delta L=&-2h\int_{\Omega} \bm{A}_t\cdot \delta\bm{u}_{mt}\,dA-2h\int_{\Omega} A_3\cdot \delta u_{m3}\,dA+\int_{\partial\Omega_q}\int_{0}^{2h}(\bm{S}^T\bm{N}-\bm{q})\cdot\delta\bm{u}(s,Z)\,da\\
&+\int_{\partial\Omega} T_{,s} \delta u_{m3}\,ds-\int_{\partial\Omega} T_{,s}\delta u_{m3}\,ds+ \int_{\partial\Omega} M\delta \alpha_m\,ds-\int_{\partial\Omega} M\delta \alpha_m\,ds=0.
\end{split}
\end{align}

Next, substituting the expressions of $\bm{A}_t$ and $A_3$ according to the shell equations \eqref{eq:ffinal12} and \eqref{eq:ffinal3} into the above equation and then doing integration by parts by Stokes' theorem, we obtain, after dropping $O(h^4,h^3k)$ terms, 
\begin{align}\label{eq:weakform}
\begin{split}
&2h\int_{\Omega}(\tr(\bm{\overline{S}}_t\nabla\delta\bm{u}_{mt})+k^\alpha_\beta\overline{S}^{\beta3}\bm{g}_\alpha\cdot\delta\bm{u}_{mt}+(\rho\ddot{\overline{\bm{x}}}_t-\overline{\bm{q}}_t)\cdot\delta\bm{u}_{mt})\,dA+2h\int_{\Omega}\big((\overline{\bm{S}_\star}\bm{n}-\overline{\bm{S}^T_{\star}}\bm{n})\cdot\nabla\delta u_{m3}\\
&-\tr(\bm{k}\overline{\bm{S}}_t)\delta u_{m3}+\frac{1}{3}h^2\nabla\cdot((\bm{S}^{(1)}_t\bm{\tau}-\bm{S}^{(1)}[\bm{x}^{(1)}\times \bm{\nu}])\delta u_{m3,s}) -\frac{1}{6}h^2\nabla\cdot (\bm{S}^{(0)}[\bm{x}^{(2)}\times \bm{\nu}]
\delta u_{m3,s})\\
&+\frac{1}{3}h^2\nabla\cdot(\bm{S}^{(1)}_t \bm{\nu}\delta u_{m3,\nu}
-\bm{S}^{(1)}[\bm{\tau}\times\bm{x}^{(1)}]\delta\alpha_{m\star})
-\frac{1}{6}h^2\nabla\cdot(\bm{S}^{(0)}[\bm{\tau}\times\bm{x}^{(2)}]\delta\alpha_{m\star})-\frac{1}{3}h^2\tr(\bm{S}^{(1)}_t\nabla\nabla\delta u_{m3})\\
&-\frac{1}{3}h^2(\rho\ddot{\bm{x}}^{1}_t-\bm{q}^{(1)}_{bt})\cdot\nabla\delta u_{m3}+\bm{m}_t\cdot\nabla\delta u_{m3}-h\bm{k}\bm{q}^-_t\cdot\nabla \delta u_{m3}+(\rho\ddot{\overline{{x}}}_3-\overline{{q}}_3)\delta u_{m3}\big)\,dA\\
=&2h\int_{\partial \Omega} \overline{\bm{S}}^T_t\bm{\nu}\cdot\delta\bm{u}_{mt}\,ds+2h\int_{\partial\Omega} \big((\overline{\bm{S}_\star}\bm{n}-\overline{\bm{S}_\star^T}\bm{n})\cdot\bm{\nu}+\frac{1}{3}h^2(\bm{1}\nabla\cdot\bm{S}^{(1)}_t-\rho\ddot{\bm{x}}^{(1)}_t+\bm{q}_{bt}^{(1)})\cdot\bm{\nu}\\
& -\frac{1}{3}h^2(\bm{S}^{(1)T}[\bm{\nu},\bm{\nu}\times\bm{x}^{(1)}])_{,s}-\frac{1}{6}h^2(\bm{S}^{(0)T}[\bm{\nu},\bm{\nu}\times\bm{x}^{(2)}])_{,s}+\bm{m}_t\cdot\bm{\nu}-h\bm{k}\bm{q}^-_t \cdot\bm{\nu}\big) \delta u_{m3}\, ds\\
&-(\frac{2}{3}h^3\int_{\partial\Omega}\bm{S}^{(1)T}[\bm{\nu},\bm{\tau}\times\bm{x}^{(1)}]\delta\alpha_{m\star}\,ds
+\frac{1}{3}h^3\int_{\partial\Omega}\bm{S}^{(0)T}[\bm{\nu},\bm{\tau}\times\bm{x}^{(2)}]\delta\alpha_{m\star}\,ds)\\
&-\int_{\partial\Omega_q}\int_{0}^{2h}(\bm{S}^T\bm{N}-\bm{q})\cdot\delta\bm{u}(s,Z)\,da,
\end{split}
\end{align}
where $\alpha_{m\star}=\arctan({u_{m3,\nu}}/({1+\bm{1}\nabla\bm{u}_{mt}[\bm{\nu},\bm{\nu}]}))\pm p \pi$. Also, we have used the decomposition $\nabla\delta u_{m3}= \delta u_{m3,s}\bm{\tau}+ \delta u_{m3,\nu}\bm{\nu}$ and have transformed the integrals $\int_{\partial\Omega} T_{,s}\delta u_{m3}\,ds$ and $\int_{\partial\Omega_q} M\delta \alpha_m\,ds$ into integrals over $\Omega$ by Stokes' theorem.

\begin{remark}
	In \eqref{eq:weakform}, the reason that the $O(h^3k)$ terms can be dropped is because they are either relatively $O(h^2)$ smaller than $2h\int_{\Omega}\tr(\bm{k}\overline{\bm{S}}_t)\delta u_{m3}\,dA$ or relatively $O(h)$ smaller than $2h\int_{\Omega}\tr(\overline{\bm{S}}_t\nabla\delta \bm{u}_{mt})\,dA$ (since $|hk|<1$). Thus in the subsequent derivations, any $O(h^3k)$ term will be put into the reminder, which are droppable for the same reasoning. We also point out that, in order to make the above decomposition of  $\nabla u_{m3}$ as well as the 2D divergence of $T_{,s}$ and $M$ well-defined, the unit vectors $\bm{\tau}$ and $\bm{\nu}$ have to be defined in $\Omega$, which can be done as follows. The boundary $\partial\Omega$ can be described  by an implicit function $F(\theta^\alpha)=0$ after eliminating the arc length variable. Then at the point in $\Omega$ with  $\theta^\alpha=\theta_0^\alpha$, $\bm{\tau}$ can be defined as the unit tangent vector of the curve $F(\theta^\alpha)=F(\theta_0^\alpha)$ at the point and $\bm{\bm{\nu}}$ can then be defined via the formula $\bm{\nu}=\bm{\tau}\wedge \bm{n}$. Note that   the variables $\theta_0^\alpha$ of $\bm{\tau}$ and $\bm{\nu}$ are changed into  $\theta^\alpha$ in \eqref{eq:weakform}.
\end{remark}

Now, we are ready to address the boundary conditions, which should come from the last 3D edge term. For the 3D case, the vanishing of this term for any $\delta \bm{u}$ leads to the 3D boundary condition $\eqref{eq:lateralq}$ for arbitrary $Z$, which, obviously, a 2D shell theory cannot satisfy. So, for a 2D shell theory one needs to make some special choices for $\delta \bm{u}$. Here, the criterion is that the lateral force $\bm{q}$ should generate the virtual work; at the same time for such a choice, the remaining three terms on the right-hand side should give the virtual work done by the external 3D force at the edge so that after the vanishing of the last term, \eqref{eq:weakform}  gives the 2D shell virtual work principle (that is the main reason that the above  calculations are about). According to this criterion, we choose
\begin{align}
\begin{split}
\delta\bm{u}(s,Z)=&\delta\bm{u}_{mt}+\delta u_{m3}\bm{n}+(Z-h)( \delta u_{m3,s}(\bm{\nu}\times \bm{x}^{(1)})-\delta\alpha_{m}(\bm{\tau}\times\bm{x}^{(1)}))\\
&+\frac{1}{2}(Z^2-h^2)(\delta u_{m3,s}(\bm{\nu}\times \bm{x}^{(2)})-\delta\alpha_{m}(\bm{\tau}\times\bm{x}^{(2)}))
\end{split}
\end{align}
on $\partial\Omega_q$. Then the vanishing of the last integral of \eqref{eq:weakform} leads to
\begin{align}\label{eq:bc}
\begin{split}
&\int_{\partial \Omega_q}\int_{0}^{2h} \bm{S}^{T}_t\bm{N}\cdot \delta\bm{u}_{mt}\,da+\int_{\partial\Omega_q} \big(\int_{0}^{2h} \bm{S}^{T}\bm{N}\cdot\bm{n}\sqrt{g_\tau}\,dZ-(\int_{0}^{2h} (Z-h)\bm{S}^T\bm{N}\cdot (\bm{\nu}\times\bm{x}^{(1)})\sqrt{g_\tau}\,dZ)_{,s}\\
&-(\int_{0}^{2h} \frac{1}{2}(Z^2-h^2)\bm{S}^T\bm{N}\cdot (\bm{\nu}\times\bm{x}^{(2)})\sqrt{g_\tau}\,dZ)_{,s}\big)\delta u_{m3}\,ds-[\int_{\partial\Omega_q}\int_{0}^{2h} (Z-h)\bm{S}^T\bm{N}\cdot (\bm{\tau}\times\bm{x}^{(1)})\delta\alpha_m\,da\\
&+\int_{\partial\Omega_q}\int_{0}^{2h} \frac{1}{2}(Z^2-h^2)\bm{S}^T\bm{N}\cdot (\bm{\tau}\times\bm{x}^{(2)})\delta\alpha_m\,da]\\
=&\int_{\partial \Omega_q}\int_{0}^{2h} \bm{q}_t\cdot \delta\bm{u}_{mt}\,da+\int_{\partial\Omega_q} \big(\int_{0}^{2h} q_3\sqrt{g_\tau}\,dZ-(\int_{0}^{2h} (Z-h)\bm{q}\cdot (\bm{\nu}\times\bm{x}^{(1)})\sqrt{g_\tau}\,dZ)_{,s}\\
&-(\int_{0}^{2h} \frac{1}{2}(Z^2-h^2)\bm{q}\cdot (\bm{\nu}\times\bm{x}^{(2)})\sqrt{g_\tau}\,dZ)_{,s}\big)\delta u_{m3}\,ds-[\int_{\partial\Omega_q}\int_{0}^{2h} (Z-h)\bm{q}\cdot (\bm{\tau}\times\bm{x}^{(1)})\delta\alpha_m\,da \\
&+\int_{\partial\Omega_q}\int_{0}^{2h}\frac{1}{2}(Z^2-h^2)\bm{q}\cdot (\bm{\tau}\times\bm{x}^{(2)})\delta\alpha_m\,da]
\end{split}
\end{align}
Next we shall examine each integral on the left-hand side of \eqref{eq:bc} upon using the Taylor expansions (i.e., specializing to the 2D shell theory) and its counterpart on the right-hand side.

1. The first integral $L_1$on the left-hand side of \eqref{eq:bc} is found to be 
\begin{align}
\begin{split}
L_1= 2h\int_{\partial\Omega_q} \overline{\bm{S}}^T_t\bm{\nu}\cdot\delta\bm{u}_{mt}\,ds+O(h^3),\\
\end{split}
\end{align}
which agrees with the first integral on the right-hand side of \eqref{eq:weakform} over $\partial\Omega_q$. 

The applied in-plane force per unit arc length of $\partial\Omega_q$ is $\widehat{\bm{q}}_t=\int_{0}^{2h} \bm{q}_t\sqrt{g_\tau}\,dZ$, so the first integral $R_1$ on the right-hand side of \eqref{eq:bc} can be written as
\begin{align}
R_1=\int_{\partial \Omega_q}(\int_{0}^{2h} \bm{q}_t\sqrt{g_\tau}\,dZ)\cdot \delta\bm{u}_{mt}\,ds=\int_{\partial \Omega_q}\widehat{\bm{q}}_t \cdot \delta\bm{u}_{mt}\,ds,
\end{align}
which is the virtual work by the applied 3D  in-plane force.

2. The second integral $L_2$ on the left-hand side of \eqref{eq:bc} is
\begin{align}
\begin{split}
L_2= & 2h\int_{\partial\Omega_q} \big((\overline{\bm{S}_\star}\bm{n}-\overline{\bm{S}_\star^T}\bm{n})\cdot\bm{\nu}+\frac{1}{3}h^2(\bm{1}\nabla\cdot\bm{S}^{(1)}_t-\rho\ddot{\bm{x}}^{(1)}_t+\bm{q}_{bt}^{(1)})\cdot\bm{\nu}-\frac{1}{3}h^2(\bm{S}^{(1)T}[\bm{\nu},\bm{\nu}\times\bm{x}^{(1)}])_{,s}\\
&-\frac{1}{6}h^2(\bm{S}^{(0)T}[\bm{\nu},\bm{\nu}\times\bm{x}^{(2)}])_{,s} +\bm{m}_t\cdot\bm{\nu}-h\bm{k}\bm{q}^-_t \cdot\bm{\nu}\big) \delta u_{m3}\, ds+O(h^4,h^3k),
\end{split}
\end{align}
where use has been made of \eqref{eq:mmm} and \eqref{eq:keq2}. We see that $L_2$ is same as the second integral on the right-hand side of \eqref{eq:weakform} over $\partial \Omega_q$.

The applied shear force per unit arc length of $\partial\Omega_q$ is $q_{s3}=\int_{0}^{2h} q_3\sqrt{g_\tau}\,dZ$.
The twisting moment  at the edge about the middle surface due to the applied force $\bm{q}$ is written as
\begin{align}
\begin{split}
T_q&=\int_{0}^{2h} ((\bm{x}-\bm{x}(\bm{r},h))\times \bm{q})\cdot \bm{N}_m\sqrt{g_\tau}\,dZ\\
&= \int_{0}^{2h} (Z-h)(\bm{\nu}\times \bm{x}^{(1)})\cdot\bm{q}\sqrt{g_\tau}\,dZ
+\int_{0}^{2h}\frac{1}{2}(Z^2-h^2)(\bm{\nu}\times \bm{x}^{(2)})\cdot\bm{q}\sqrt{g_\tau}\,dZ+O(h^4,h^3k),
\end{split}
\end{align}
whose derivative $T_{q,s}$ with respect to the arc length variable is equivalent to a downward shear force. Then, the second integral $R_2$ on the right-hand side is
\begin{align}
\begin{split}
R_2= \int_{\partial\Omega_q} (q_{s3}-T_{q,s})\delta u_{m3}\,ds+O(h^4,h^3k)=\int_{\partial\Omega_q}\widehat{q}_3\delta u_{m3}\,ds+O(h^4,h^3k),
\end{split}
\end{align}
where $\widehat{q}_3$ is the total effective applied shear force per unit arc length of $\partial\Omega_q$, and one can see $R_2$ is the virtual work done by the applied 3D force due to the virtual displacement $\delta u_{m3}$.

3. The third term $L_3$ on the left-hand side of \eqref{eq:bc} is
\begin{align}\label{eq:L3}
\begin{split}
L_3
=-\frac{2}{3}h^3\int_{\partial\Omega_q}\bm{S}^{(1)T}[\bm{\nu},\bm{\tau}\times\bm{x}^{(1)}]\delta\alpha_{m\star}\,ds-\frac{1}{3}h^3\int_{\partial\Omega_q}\bm{S}^{(0)T}[\bm{\nu},\bm{\tau}\times\bm{x}^{(2)}]\delta\alpha_{m\star}\,ds+O(h^4,h^3k),
\end{split}
\end{align}
which is the same as the third term on the right-hand side of \eqref{eq:weakform} over $\partial \Omega_q$.

The bending moment at the edge point about the middle surface due to the applied force $\bm{q}$ is
\begin{align}
\begin{split}
\widehat{m}_3&=\int_{0}^{2h} ((\bm{x}-\bm{x}(\bm{r},h)\times \bm{q})\cdot \bm{T}_m\sqrt{g}_\tau\,dZ\\
&=\int_{0}^{2h} (Z-h) (\bm{\tau}\times \bm{x}^{(1)})\cdot\bm{q}\sqrt{g}_\tau\,dZ
+\int_{0}^{2h}\frac{1}{2} (Z^2-h^2) (\bm{\tau}\times \bm{x}^{(2)})\cdot\bm{q}\sqrt{g}_\tau\,dZ+O(h^4,h^3k).
\end{split}
\end{align}
Then, the third term $R_3$ on the right-hand side of \eqref{eq:bc} can be written as
\begin{align}
\begin{split}
R_3=-\int_{\partial\Omega_q}\widehat{m}_3\delta\alpha_{m\star}\,ds+(h^4,h^3k),
\end{split}
\end{align}
which is the virtual work by the applied 3D force due to the virtual rotation angle.

Finally, the equalities $L_i=R_i\ (i=1,2,3)$ lead to the following boundary conditions on the traction edge $\partial\Omega_q$:
\begin{align}
&2h \overline{\bm{S}}^T_{t}\bm{\nu}=\widehat{\bm{q}}_t,\label{eq:work12}\\
\begin{split}
&2h\big((\overline{\bm{S}_\star}\bm{n}-\overline{\bm{S}^T_\star}\bm{n})\cdot\bm{\nu}+\frac{1}{3}h^2(\bm{1}\nabla\cdot\bm{S}^{(1)}_t-\rho\ddot{\bm{x}}^{(1)}_t+\bm{q}_{bt}^{(1)})\cdot\bm{\nu}\\
&+\frac{1}{3}h^2(\bm{S}^{(1)T}_t[\bm{\nu},\bm{\tau}])_{,s}+\bm{m}_t\cdot\bm{\nu}-h\bm{k}\bm{q}^-_t \cdot\bm{\nu}\big)=\widehat{q}_3,\label{eq:work3}
\end{split}\\
&\frac{2}{3}h^3\bm{S}^{(1)T}_t[\bm{\nu},\bm{\nu}]+\frac{2}{3}h^3\bm{S}^{(1)T}[\bm{\nu},\bm{\tau}\times\bm{u}^{(1)}]+\frac{1}{3}h^3\bm{S}^{(0)T}[\bm{\nu},\bm{\tau}\times\bm{u}^{(2)}]=\widehat{m}_3,\label{eq:workbending}
\end{align}
where $\bm{q}_t$ and $\widehat{q}_3$ are respectively the applied in-plane force and total effective shear force (per unit arc length of $\partial\Omega_q$), and $\widehat{m}_3$ is the applied bending moment about the middle surface, which are all supposed to be prescribed. In the above equations, we have made use of $\bm{x}^{(1)}=\bm{n}+\bm{u}^{(1)}$ and $\bm{x}^{(2)}=\bm{u}^{(2)}$, which result from the relation $\bm{x}=\bm{X}+\bm{u}$, and  
the two terms $-\frac{1}{3}h^2(\bm{S}^{(0)T}[\bm{\nu},\bm{\nu}\times \bm{u}^{(1)}])_{,s}$ and $-\frac{1}{6}h^2(\bm{S}^{(0)T}[\bm{\nu},\bm{\nu}\times \bm{u}^{(2)}])_{,s}$ have been dropped in \eqref{eq:work3} for the following reason: for large deformations, they are $O(h^2)$ smaller than $(\overline{\bm{S}_\star}\bm{n}-\overline{\bm{S}^T_\star}\bm{n})\cdot\bm{\nu}$, while for small deformations, they are  smaller than $\frac{1}{3}h^2(\bm{S}^{(1)T}_t[\bm{\nu},\bm{\tau}])_{,s}$. Thus,  no matter for large or small deformations they can be dropped.

Based on work conjugates, on the displacement edge $\partial \Omega_0$, the boundary conditions are:
\begin{align}\label{eq:bc2}
\bm{u}_{mt}= \widehat{\bm{u}}_{mt},\quad u_{m3}=\widehat{u}_{m3},\quad \alpha_{m\star}=\widehat{\alpha}_{m}\iff \frac{u_{m3,\nu}}{1+\bm{1}\nabla \bm{u}_{mt}[\bm{\nu},\bm{\nu}]}=\tan(\widehat{\alpha}_m),
\end{align}
where $\bm{u}_m=\bm{u}^{(0)}+h\bm{u}^{(1)}+O(h^2)$, and $\widehat{\bm{u}}_m$ and $\widehat{\alpha}_m$ are respectively the prescribed displacement and rotation angle of the middle surface.

Upon using these boundary conditions for the right-hand side of \eqref{eq:weakform}, we obtain the 2D shell virtual work principle (as the right-hand side represents the virtual work done by the applied effective 3D force at the edge):
\begin{align}\label{eq:virtualweak}
\begin{split}
&2h\int_{\Omega}(\tr(\bm{\overline{S}}_t\nabla\delta\bm{u}_{mt})+k^\alpha_\beta\overline{S}^{\beta3}\bm{g}_\alpha\cdot\delta\bm{u}_{mt}+(\rho\ddot{\overline{\bm{x}}}_t-\overline{\bm{q}}_t)\cdot\delta\bm{u}_{mt})\,dA\\
&+2h\int_{\Omega}\big((\overline{\bm{S}_\star}\bm{n}-\overline{\bm{S}^T_{\star}}\bm{n})\cdot\nabla\delta u_{m3}-\tr(\bm{k}\overline{\bm{S}}_t)\delta u_{m3}-\frac{1}{3}h^2\tr(\bm{S}^{(1)}_t\nabla\nabla\delta u_{m3})\\
&-\frac{1}{3}h^2(\rho\ddot{\bm{x}}^{1}_t-\bm{q}^{(1)}_{bt})\cdot\nabla\delta u_{m3}+\bm{m}_t\cdot\nabla\delta u_{m3}-h\bm{k}\bm{q}^-_t\cdot\nabla \delta u_{m3}+(\rho\ddot{\overline{{x}}}_3-\overline{{q}}_3)\delta u_{m3}\big)\,dA\\
=&\int_{\partial \Omega_q}\widehat{\bm{q}}_t \cdot \delta\bm{u}_{mt}\,ds+\int_{\partial\Omega_q}\widehat{q}_3\delta u_{m3}\,ds- \int_{\partial\Omega_q}\widehat{m}_3\delta\alpha_{m\star}\,ds.
\end{split}
\end{align}
In obtaining the above equation, the following four terms in \eqref{eq:weakform} have been dropped:
\begin{align}
&\frac{2}{3}h^3 \nabla\cdot((\bm{S}^{(1)}_t\bm{\tau}-\bm{S}^{(1)}[\bm{x}^{(1)}\times \bm{\nu}])\delta u_{m3,s}),\quad -\frac{1}{3}h^3 \nabla\cdot(\bm{S}^{(0)}[\bm{x}^{(2)}\times \bm{\nu}]\delta u_{m3,s})\label{eq:e1}\\
&\frac{2}{3}h^3 \nabla\cdot(\bm{S}^{(1)}_t \bm{\nu}\delta u_{m3,\nu}-\bm{S}^{(1)}[\bm{\tau}\times\bm{x}^{(1)}]\delta\alpha_{m\star}),\quad-\frac{1}{3}h^3 \nabla\cdot(\bm{S}^{(0)}[\bm{\tau}\times\bm{x}^{(2)}]\delta\alpha_{m\star})\label{eq:e2}
\end{align} 
which can be justified as follows. From the relations
$\bm{x}^{(1)}=\bm{n}+\bm{u}^{(1)}$ and $\bm{x}^{(2)}=\bm{u}^{(2)}$, the two terms in \eqref{eq:e1} can be simplified as $-\frac{2}{3}h^3\nabla\cdot(\bm{S}^{(1)}\bm{u}^{(1)})\delta u_{m3,s}$ and $-\frac{1}{3}h^3 \nabla\cdot(\bm{S}^{(0)}[\bm{u}^{(2)}\times \bm{\nu}]\delta u_{m3,s})$ . From \eqref{eq:rotation angle},
the variation of $\alpha_{m\star}$ is calculated by
\begin{align}
\delta\alpha_{m\star}=\frac{(1+\bm{1}\nabla\bm{u}_{mt}[\bm{\nu},\bm{\nu}])\delta u_{m3,\nu}-u_{m3,\nu}\bm{1}\nabla\delta\bm{u}_{mt}[\bm{\nu},\bm{\nu}]} {(1+\bm{1}\nabla\bm{u}_{mt}[\bm{\nu},\bm{\nu}])^2+(u_{m3,\nu})^2}=\delta u_{m3,\nu}+O(\nabla\bm{u}_m\delta u_{m3,\nu},\nabla\bm{u}_m\nabla\delta\bm{u}_{mt}),
\end{align} 
where the second equality is for small deformations. In \eqref{eq:e2}, the terms related to $\nabla \delta \bm{u}_{mt}$  are relatively $O(h^2)$ smaller than $2h \tr(\bm{\overline{S}}_t\nabla\delta\bm{u}_{mt})$ and can thus be dropped. For the remaining terms left in \eqref{eq:e2} and the two terms in \eqref{eq:e1}, for large deformations, they are relatively $O(h^2)$ smaller than $2h (\overline{\bm{S}}\bm{n}-\overline{\bm{S}_\star^T}\bm{n})\nabla\delta u_{m3}$, while for small deformations they are of $ O(h^3 \bm{S}^{(1)}\bm{u}^{(1)}\delta u_{m3,s},h^3 \bm{S}^{(0)}\bm{u}^{(2)}\delta u_{m3,s})$ and $O(h^3 \bm{S}^{(1)}\bm{u}^{(1)}\delta u_{m3,\nu},h^3 \bm{S}^{(0)}\bm{u}^{(2)}\delta u_{m3,\nu})$, which are smaller than $-\frac{2}{3}h^3 \tr(\bm{S}^{(1)}_t \nabla\nabla\delta u_{m3})$. Thus they can be dropped no matter the deformation is large or small.

The 2D shell virtual work principle \eqref{eq:virtualweak} supplemented by boundary conditions \eqref{eq:work12}-\eqref{eq:workbending} and \eqref{eq:bc2} provides a framework for implementing finite element schemes, which will be left for future investigations.

\section{A Benchmark problem: the extension and inflation of an arterial segment}\label{sec:benchmark}

In this section, we apply the previously derived shell theory to study the extension and inflation of an arterial segment, for which the exact solution is available in \cite{haughton1979bifurcation}. We will compare the asymptotic solution obtained from the shell theory and the exact solution to show its validity.

Following \cite{holzapfel2010constitutive}, we consider an artery as a thick-walled circular cylindrical tube, which in its reference configuration has internal and external radii $A$ and $B$, respectively, and length $L$. So, its geometry may be described in terms of cylindrical polar coordinates $(R,\Theta,X)$ by
\begin{equation}
A\leq R\leq B,\quad 0\leq \Theta\leq 2\pi,\quad 0\leq X\leq L.
\end{equation}
They are related to the Cartesian coordinates $(X_1,X_2,X_3)$ by
\begin{equation}\label{eq:Cartesian}
X_1 = R\cos\Theta,\quad X_2=R\sin\Theta,\quad X_3=X.
\end{equation}
In the notation of the shell theory, we have the corresponding relations
\begin{equation}
\theta^1=\Theta,\quad \theta^2=X,\quad Z=R-A,\quad 2h=B-A.
\end{equation}
We choose the inner surface of the circular cylindrical tube as the base surface. Let $(\bm{e}_R,\bm{e}_\Theta,\bm{e}_X)$ denote the standard basis vectors of the cylindrical polar coordinates. A direct calculation using \eqref{eq:Cartesian} shows
\begin{align}
\bm{g}_1 =  A^2\bm{g}^1=A\bm{e}_{\Theta},\quad \bm{g}_2 =\bm{g}^2=\bm{e}_X,\quad \bm{g}_3  =\bm{g}^3 =\bm{e}_R=\bm{n}.
\end{align}
Thus the 2D gradient operator is given by $\nabla=\frac{1}{A}\frac{\partial }{\partial\Theta}\bm{e}_{\Theta}+\frac{\partial}{\partial X}\bm{e}_X$. The curvature tensor is calculated by $
\bm{k}=-\bm{n}_{,\alpha}\otimes \bm{g}^\alpha=-\frac{1}{A}\bm{e}_{\Theta}\otimes \bm{e}_{\Theta}$,  
which implies that $H=-\frac{1}{2A}$ and $K=0$. 

In the problem of the extension and inflation of the artery, the circular cylindrical tube is assumed to undergo an axisymmetric and uniformly extensional deformation. Thus the deformed tube is described in cylindrical polar coordinates $(r,\theta,z)$ by
\begin{equation}
a\leq r\leq b,\quad 0\leq \theta\leq 2\pi,\quad  0\leq z\leq l,
\end{equation}
where $a,b$ and $l$ are the deformed counterparts of $A, B$ and $L$ respectively and deformation is given by
\begin{equation}\label{eq:x}
r=r(R),\quad \theta=\Theta,\quad z=\lambda_z X,
\end{equation}
where $\lambda_z=l/L$ is the uniform stretch in the axial direction. Let $(\bm{e}_r,\bm{e}_{\theta},\bm{e}_z)$ denote the standard basis vectors of the cylindrical polar coordinates $(r,\theta,z)$ which actually agree with $(\bm{e}_R,\bm{e}_\Theta,\bm{e}_X)$. In cylindrical polar coordinates,  the shell equations \eqref{eq:ffinal12} and \eqref{eq:ffinal3}  take the following form
\begin{align}
&\frac{1}{A}\frac{\partial \overline{S}_{\Theta\theta}}{\partial \Theta} +\frac{\partial \overline{S}_{X\theta}}{\partial X} + \frac{1}{A}\overline{S}_{\Theta r}=\rho\ddot{\overline{x}}_{\Theta}-\frac{\mu(2h)q^+_\Theta+q^-_\Theta}{2h}-\overline{q}_{b\Theta}, \label{eq:circular1}\\
&\frac{1}{A}\frac{\partial \overline{S}_{\Theta z}}{\partial \Theta} +\frac{\partial \overline{S}_{Xz}}{\partial X}=\rho\ddot{\overline{x}}_{X}-\frac{\mu(2h)q^+_X+q^-_X}{2h}-\overline{q}_{bX}, \label{eq:circular2}\\
\begin{split}
&\frac{1}{A}(\frac{\partial \overline{S_\star}_{\Theta r}}{\partial\Theta}-\frac{\partial\overline{S^T_\star}_{\Theta r}}{\partial \Theta})+\frac{\partial\overline{S_\star}_{ Xr}}{\partial X}-\frac{\partial\overline{S^T_\star}_{ Xr}}{\partial X}-\frac{1}{A}\overline{S}_{\Theta\theta}\\
&+\frac{1}{3}h^2(\frac{1}{A^2}\frac{\partial^2 S^{(1)}_{\Theta\theta}}{\partial\Theta^2}+\frac{1}{A}\frac{\partial^2 S^{(1)}_{\Theta z}}{\partial\Theta\partial X}+\frac{1}{A}\frac{\partial^2 S^{(1)}_{X\theta}}{\partial\Theta\partial X}+\frac{\partial^2 S^{(1)}_{Xz}}{\partial X^2})\\
=&\rho\ddot{\overline{x}}_{R}-\frac{\mu(2h)q^+_R+q^-_R}{2h}-\overline{q}_{bR}+\frac{1}{3}h^2(\frac{1}{A}\frac{\partial}{\partial\Theta}(\rho\ddot{{x}}^{(1)}_{\Theta}-q^{(1)}_{b\Theta})+\frac{\partial }{\partial X}(\rho\ddot{x}^{(1)}_{X}-q^{(1)}_{bX}))\\
& -(\frac{1}{A}\frac{\partial m_\Theta}{\partial \Theta}+\frac{\partial m_X}{\partial X})-\frac{h}{A^2}\frac{\partial q^-_{\Theta}}{\partial\Theta}, \label{eq:circular3}
\end{split}
\end{align}
where $\overline{\bm{S}}, \overline{\bm{S}_{\star}}, \overline{\bm{S}^T_{\star}}, \overline{\bm{x}}$ and $\overline{\bm{q}}_b$ are defined below \eqref{eq:ffinal3}.  

The deformation gradient arsing from the deformation  \eqref{eq:x} is given by
\begin{equation}\label{eq:deformation}
\bm{F}=\frac{r}{R}\bm{e}_{\theta}\otimes\bm{e}_{\Theta} + \lambda_{z}\bm{e}_z\otimes\bm{e}_X+r'\bm{e}_r\otimes\bm{e}_R.
\end{equation} 
On the inner and outer surfaces of the circular cylindrical tube, we consider the traction boundary conditions caused by the internal pressure $P$
\begin{align}\label{eq:traction}
&\bm{q}^- =P\bm{F}^{(0)-T}\bm{n}=P\frac{\lambda_z a}{A}\bm{e}_R, \quad \bm{q}^+=0.
\end{align}
On its end surface, we impose a resultant axial force 
\begin{equation}\label{eq:lateral}
F=2\pi \int_A^B S_{Xz}R\,dR-\pi a^2 P.
\end{equation}

The artery is modelled as an incompressible hyperelastic material reinforced by two symmetrically disposed families of fibres, which has a strain energy function \cite{holzapfel2000new} given by 
\begin{equation}\label{eq:strain energy}
W(I_1,I_4,I_6)=\frac{c}{2}(I_1-3)+\frac{k_1}{2k_2}\sum_{i=4,6}(e^{k_2(I_i-1)^2}-1),
\end{equation}
where $I_1=\tr(\bm{C})$ is the first principal invariant of the right Cauchy-Green tensor $\bm{C}=\bm{F}^T\bm{F}$, and $I_4=\bm{M}\cdot (\bm{C}\bm{M})$ and $I_6=\bm{M}'\cdot (\bm{C}\bm{M}')$, where
the unit vectors $\bm{M}=\cos\varphi\bm{e}_\Theta+\sin\varphi\bm{e}_X$ and $\bm{M}'=-\cos\varphi\bm{e}_\Theta+\sin\varphi\bm{e}_X$
represent the directions of the two fibres. It follows from \eqref{eq:deformation}  that $I_4$ and $I_6$ are 
\begin{equation}
I_4=I_6=\frac{r^2}{R^2}\cos^2\varphi +\lambda_z^2\sin^2\varphi:=I.
\end{equation}
For the strain energy function \eqref{eq:strain energy},  the associated nominal stress is given by
\begin{equation}\label{eq:nomial}
\bm{S} = c \bm{F}^T + 2k_1(I_4-1)e^{k_2(I_4-1)^2}\bm{M}\otimes \bm{F}\bm{M} + 2k_1(I_6-1)e^{k_2(I_6-1)^2}\bm{M}'\otimes \bm{F}\bm{M}'-p\bm{F}^{-1}.\\
\end{equation}

First, substituting \eqref{eq:deformation} into \eqref{eq:nomial} and doing a Taylor expansion yield
\begin{align}
\begin{split}
&\bm{S}^{(0)}=(c\frac{r_0}{A}-p_0\frac{A}{r_0}+4k_1(I_0-1)e^{k_2(I_0-1)^2}\frac{r_0}{A}\cos^2\varphi)\bm{e}_{\Theta}\otimes\bm{e}_{\theta}\\
&\qquad\quad+(c\lambda_z-\frac{p_0}{\lambda_z}+4k_1(I_0-1)e^{k_2(I_0-1)^2}\lambda_z\sin^2\varphi)\bm{e}_X\otimes\bm{e}_z\\
&\qquad\quad+(cr_1-\frac{p_0}{r_1})\bm{e}_{R}\otimes\bm{e}_{r},
\end{split}\\
\begin{split}
&\bm{S}^{(1)}=(c\frac{r_1A-r_0}{A^2}-p_0\frac{r_0-r_1A}{r_0^2}-p_1\frac{A}{r_0}+4k_1((1+2k_2(I_0-1)^2)I_1\frac{r_0}{A}\\
&\qquad\quad+(I_0-1)\frac{r_1A-r_0}{A^2})e^{k_2(I_0-1)^2}\cos^2\varphi)\bm{e}_{\Theta}\otimes\bm{e}_{\theta}\\
&\qquad\quad+(-\frac{p_1}{\lambda_z}+4k_1(1+2k_2(I_0-1)^2)e^{k_2(I_0-1)^2}I_1\lambda_z\sin^2\varphi)\bm{e}_X\otimes\bm{e}_z\\
&\qquad\quad +(cr_2-\frac{p_1}{r_1}+\frac{p_0r_2}{r_1^2})\bm{e}_R\otimes\bm{e}_r,
\end{split}
\end{align}
where $r_i, p_i, I_i$ denote the $i$th derivatives of $r, p, I$ with respect to $Z$ at $Z=0$, respectively; in particular, we have 
\begin{equation}
I_0=I|_{Z=0}=\frac{r_0^2}{A^2}\cos^2\varphi+\lambda_z^2\sin^2\varphi,\quad I_1=\frac{\partial I}{\partial Z}|_{Z=0}=\frac{2r_0(r_1A-r_0)}{A^3}\cos^2\varphi.
\end{equation}

Next we obtain from \eqref{eq:S0} and \eqref{eq:coneq1} the recurrence relation for $p_0$ and $r_1$:
\begin{equation}\label{eq:recurrence1}
p_0  = c\frac{A^2}{\lambda_z^2 r_0^2}+P,\quad r_1  =\frac{A}{\lambda_z r_0},\quad 
\end{equation}
and from \eqref{eq:p1} and \eqref{eq:x2} the recurrence relation for $p_1$ and $r_2$:
\begin{equation}\label{eq:recurrence2}
p_1  = -c\frac{(\lambda_z r_0^2-A^2)^2}{\lambda_z^3 Ar_0^4}-4k_1e^{k_2(I_0-1)^2}\frac{(I_0-1) \cos^2\varphi}{\lambda_z A},\quad r_2  =\frac{\lambda_z r_0^2-A^2}{\lambda_z^2 r_0^3}.
\end{equation}

Finally the only nontrivial shell equation \eqref{eq:circular3} becomes 
\begin{equation}\label{eq:final1}
\frac{1}{A}(S^{(0)}_{\Theta\theta}+hS^{(1)}_{\Theta\theta})=\frac{q^-_{R}}{2h}=\frac{P}{2h}\frac{\lambda_z r_0}{A}.
\end{equation}
Substituting the recurrence relations \eqref{eq:recurrence1} and \eqref{eq:recurrence2} into the above equation, we obtain an equation involving $r_0$ only as expected
\begin{align}\label{eq:final2}
\begin{split}
&\varrho-c{\lambda_a^{-4}\lambda_z^{-3}}(\lambda_a^4\lambda_z^2-1)-4k_1e^{k_2(I_0-1)^2}(I_0-1){\lambda_z}^{-1}\cos^2\varphi+h^*\big(\varrho\lambda_a^{-2}\lambda_z^{-1}\\
&+c\frac{1}{2}\lambda_a^{-6}\lambda_z^{-4}(\lambda_a^6\lambda_z^3-2\lambda_a^4\lambda_z^2+3\lambda_a^2\lambda_z-2)+2k_1e^{k_2(I_0-1)^2}\lambda_a^{-2}\lambda_z^{-2}\cos^2\varphi \\
&\times ((\lambda_a^2\lambda_z-2)(I_0-1)+2\lambda_a^2(\lambda_a^2\lambda_z-1)(1+2k_2(I_0-1)^2)\cos^2\varphi)\big)\\
&+h^{*2}\frac{1}{2}\varrho\lambda_a^{-4}\lambda_z^{-2}(\lambda_a^2\lambda_z-1)=0,
\end{split}
\end{align}
where the scales are set as $h^*=2h/A$, $P=\varrho 2h/A$ and  $\lambda_a=r_0/A=a/A$.  We observe from \eqref{eq:final2} that
\begin{align}
\varrho=c{\lambda_a^{-4}\lambda_z^{-3}}(\lambda_a^4\lambda_z^2-1)+4k_1(I_0-1)e^{k_2(I_0-1)^2}{\lambda_z}^{-1}\cos^2\varphi+O(h^*).
\end{align}
Substituting the above equation into the $O(h^*)$ term of \eqref{eq:final2}, we have
\begin{align}
\begin{split}\label{eq:rho}
P=&\varrho h^*=h^*(c{\lambda_a^{-4}\lambda_z^{-3}}(\lambda_a^4\lambda_z^2-1)+4k_1(I_0-1)e^{k_2(I_0-1)^2}{\lambda_z}^{-1}\cos^2\varphi)\\
&-h^{*2}\big(\frac{1}{2}c\lambda_a^{-6}\lambda_z^{-4}(\lambda_a^6\lambda_z^3+3\lambda_a^2\lambda_z-4)+2k_1e^{k_2(I_0-1)^2}\lambda_a^{-2}\lambda_z^{-2}\cos^2\varphi \\
&\times (\lambda_a^2\lambda_z(I_0-1)+2\lambda_a^2(\lambda_a^2\lambda_z-1)(1+2k_2(I_0-1)^2)\cos^2\varphi)\big) +O(h^{*3}).
\end{split}
\end{align}
Then according to \eqref{eq:work12}, the boundary condition \eqref{eq:lateral} gives
\begin{align}
2h((1+\frac{h}{A}){S}^{(0)}_{Xz}+h{S}^{(1)}_{Xz})=\frac{F+\pi a^2 P}{2\pi A}.
\end{align}
Substituting \eqref{eq:rho} into above equation, we have
\begin{align}
\begin{split}\label{eq:FF}
F^*=&h^*(c\lambda_a^{-2}\lambda_z^{-3}(2\lambda_a^2\lambda_z^4-\lambda_a^4\lambda_z^2-1)+4k_1e^{k_2(I_0-1)^2}\lambda_z^{-1}(I_0-1)(2\lambda_z^2\sin^2\varphi-\lambda_a^2\cos^2\varphi))\\
&+h^{*2}\big(\frac{1}{2}c\lambda_a^{-4}\lambda_z^{-4}(\lambda_a^6\lambda_z^3+2\lambda_a^4\lambda_z^5-2\lambda_a^4\lambda_z^2-3\lambda_a^2\lambda_z+2)\\
&+2k_1e^{k_2(I_0-1)^2} \lambda_z^{-2}\big((I_0-1)((\lambda_a^2\lambda_z-2)\cos^2\varphi+2\lambda_z^3\sin^2\varphi)\\
&+2(\lambda_a^2\lambda_z-1)(1+2k_2(I_0-1)^2)(\lambda_a^2\cos^4\varphi-2\lambda_z^2\sin^2\varphi\cos^2\varphi)\big)\big)+O(h^{*3}),
\end{split}
\end{align}
where $F^*=F/(\pi A^2)$ is the normalized resultant axial force. Equations \eqref{eq:rho} and \eqref{eq:FF} form the asymptotic solution of the problem.

On the other hand, the problem has an exact solution of the following form \cite{haughton1979bifurcation}:
\begin{align}
&P =\int_{\lambda_b}^{\lambda_a} (\lambda^2\lambda_z-1)^{-1}\psi_\lambda\,d\lambda,\label{eq:lambda}\\
&F  = \pi A^2(\lambda_a^2\lambda_z-1)\int_{\lambda_b}^{\lambda_a}(\lambda^2\lambda_z-1)^{-2}(2\lambda_z\psi_{\lambda_z} -\lambda\psi_\lambda)\lambda\,d\lambda,\label{eq:lambdaz}
\end{align}
where $\lambda_b=b/B=\sqrt{\lambda_z^{-1}((\lambda_a^2\lambda_z-1)A^2/B^2+1)}$, $\psi_{\lambda}={\partial\psi}/{\partial \lambda}$, $\psi_{\lambda_z}={\partial \psi}/{\partial \lambda_z}$, and $\psi$ is given by
\begin{equation}\label{eq:psi}
\psi(\lambda,\lambda_z)=\frac{c}{2}(\lambda^2+\lambda_z^2+\lambda^{-2}\lambda_z^{-2}-3) +\frac{k_1}{k_2}(e^{k_2(\lambda^2\cos^2\varphi+\lambda_z^2\sin^2\varphi-1)^2}-1),
\end{equation}
Doing a routine Taylor expansion, we see that
\begin{align}
&P=h^*\lambda^{-1}_a\lambda_z^{-1}{\psi_{\lambda}(\lambda_a,\lambda_z)}-h^{*2} \frac{1}{2}\lambda_a^{-3}\lambda_z^{-2}(\psi_{\lambda}(\lambda_a,\lambda_z) +\lambda_a(\lambda_a^2\lambda_z-1)\psi_{\lambda\lambda}(\lambda_a,\lambda_z))+O(h^{*3}),\label{eq:exact1}\\
\begin{split}\label{eq:exact2}
&F^*=h^* \lambda_z^{-1}(2\lambda_z\psi_{\lambda_z}(\lambda_a,\lambda_z)-\lambda_a\psi_{\lambda}(\lambda_a,\lambda_z))+h^{*2}\frac{1}{2}\lambda_a^{-1}\lambda_z^{-2}\big(2\lambda_a\lambda_z^2\psi_{\lambda_z}(\lambda_a,\lambda_z)-\psi_\lambda(\lambda_a,\lambda_z)\\
&\quad\quad\ +(\lambda_a^2\lambda_z-1)(\lambda_a\psi_{\lambda\lambda}(\lambda_a,\lambda_z)-2\lambda_z\psi_{\lambda\lambda_z}(\lambda_a,\lambda_z))\big)+O(h^{*3}),
\end{split}
\end{align}
where $\psi_{\lambda\lambda}={\partial^2\psi}/{\partial\lambda^2}$ and $\psi_{\lambda\lambda_z}={\partial^2\psi}/{\partial\lambda\partial\lambda_z}$.
If the expansions are carried out on the middle surface, then the $O(h^{*2})$ terms are not present, and the errors are of $O(h^{*3})$ as well; see equations (6.5) and (6.6) in \cite{fu2016localized}. Using \eqref{eq:psi}, it is easy to check that the exact solution \eqref{eq:exact1} and \eqref{eq:exact2} are the same as the asymptotic solution \eqref{eq:rho} and \eqref{eq:FF}, validating the shell equations.

To illustrate a numerical example, we set the geometrical and material parameters of the artery as in Table \ref{tab:1}; these parameters are cited from \cite{holzapfel2000new} and are given for a carotid artery from a rabbit.

\begin{figure}[!h]
	\centering
	\includegraphics[width=0.39\linewidth]{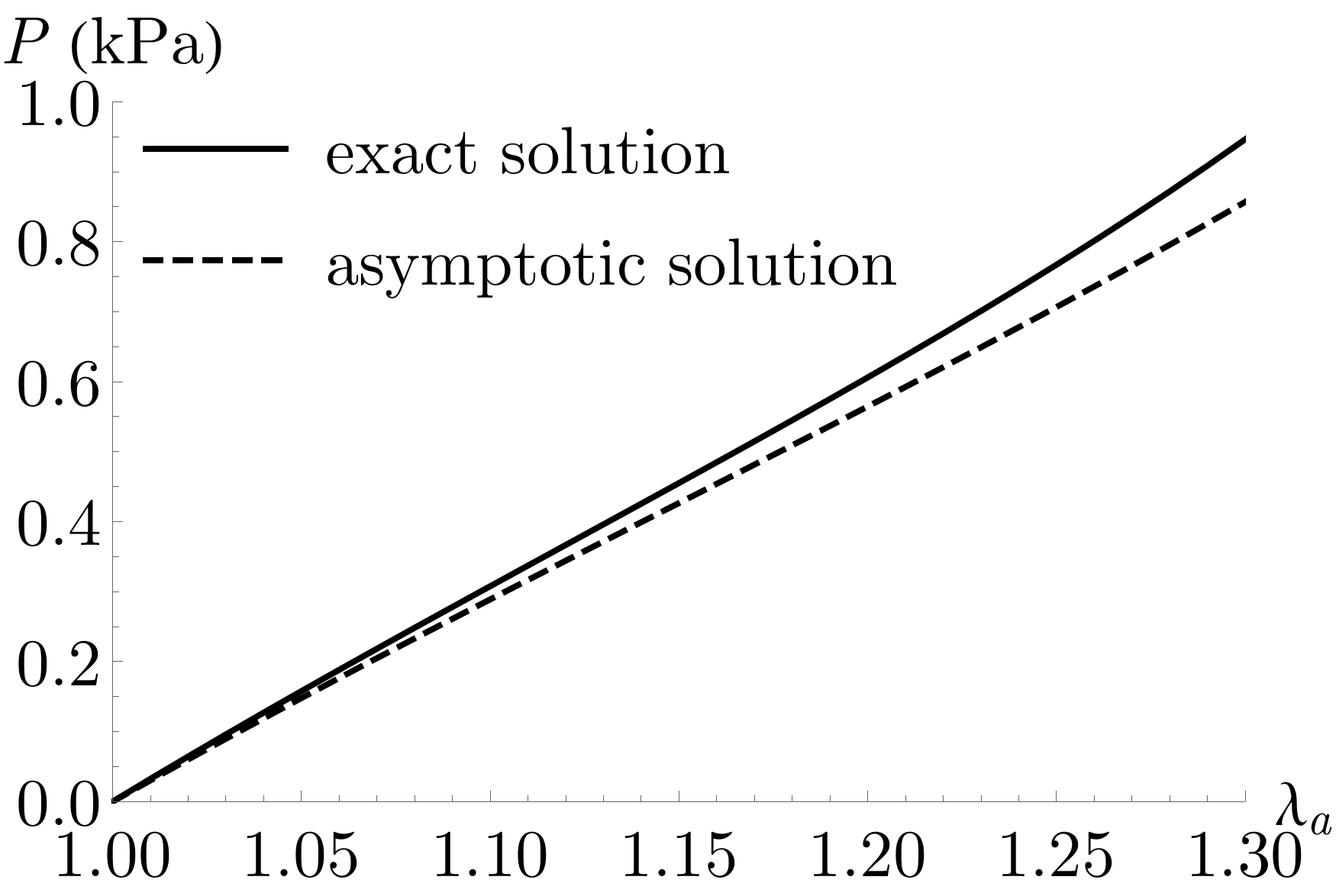}(a)\qquad\qquad
	\includegraphics[width=0.39\linewidth]{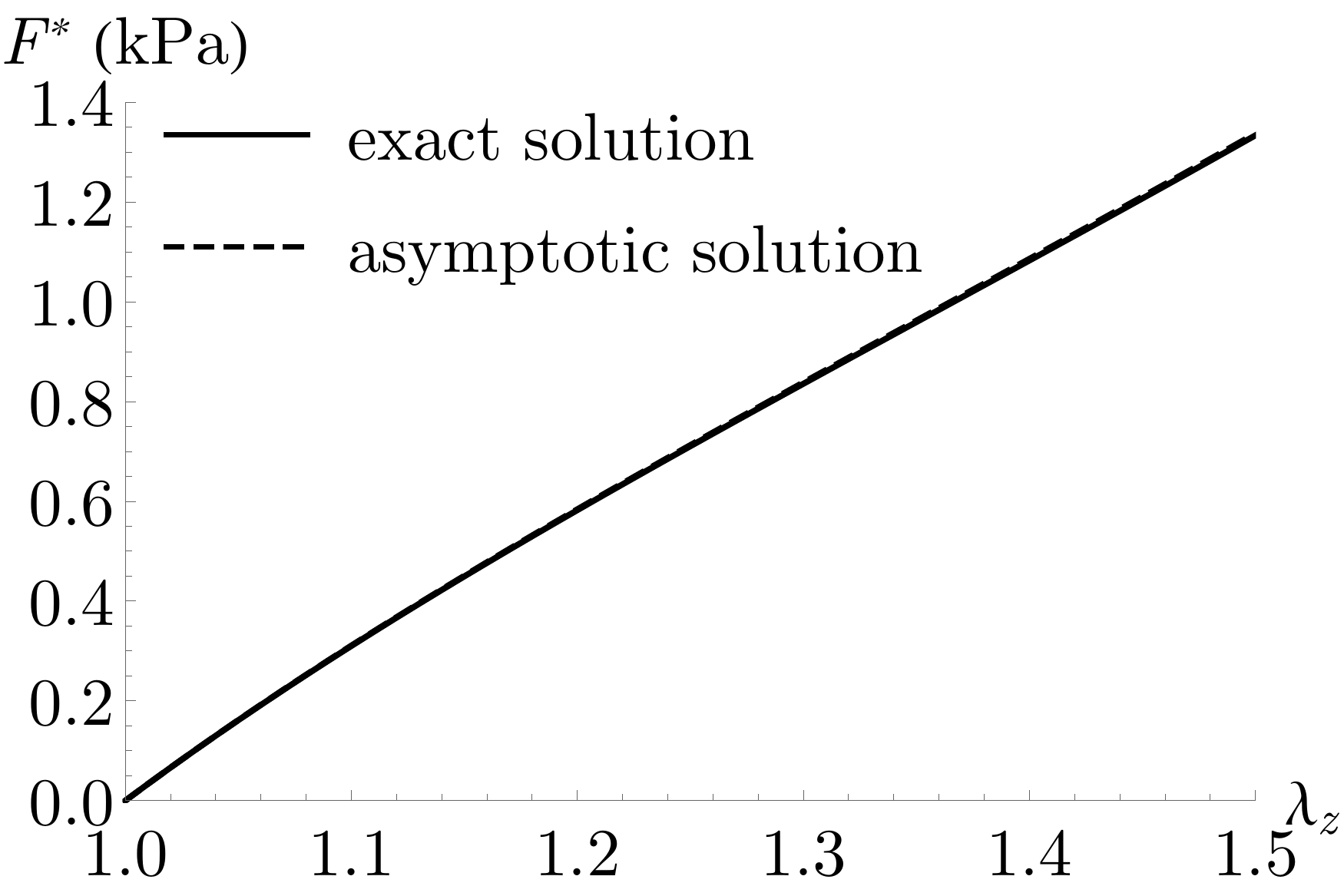}(b)
	\caption{Comparison of the exact solution and the asymptotic solution  (a) Variation of the inner pressure $P$ with respect to $\lambda_a$ for fixed $\lambda_z=1$ (b) Variation of the normalized axial force $F^*=F/(\pi A^2)$ with respect to $\lambda_z$ for fixed $\lambda_a=1$}
	\label{fig:comparison}
\end{figure}

In Figure \ref{fig:comparison}, we compare the exact solution and the asymptotic solution of the pressure $P$ and the normalized resultant axial force $F^*$  for the artery described by the above parameters. It is  seen that the asymptotic solution is very close to the exact one, which can be viewed as a numerical validation of the shell equations.
\begin{table}[h!]
	\centering
	\caption{Geometrical and material data for a carotid artery from a rabbit}
	\begin{tabular}{c c c c c c c}
		\hline
		$A$ (mm) & $2h$ (mm) & $c$ (kPa) & $k_1$ (kPa) & $k_2$ (-) & $\varphi$  & $\rho$ ($\text{g}/\text{cm}^3$)\\
		\hline
		$1.43$ & $0.26$ & $3$ & $2.3632$ & $0.8393$ & $29^\circ$ & $1.19$\\
		\hline
	\end{tabular}
	\label{tab:1}
\end{table}

\section{An application: plane-strain vibrations of a pressurized artery}\label{sec:plane-strain}

As an application of the derived refined shell theory, we consider the plane-strain vibrations of an artery superimposed on a pressurized state considered in the previous section. The results may be useful in determining the material parameters of an artery. Due to the space limit, other vibration modes together with wave propagation will be reported in a separate paper. The shell equations are three nonlinear PDEs for $\bm{x}^{(0)}$. For deformations superimposed on a known state (base state), we write
$
\bm{x}^{(0)}=\bm{x}_b^{(0)} + \delta\bm{u}^{(0)}, \label{eq:6.1}$
where the known vector $\bm{x}_b^{(0)}$ is the position vector of the deformed bottom surface in the base state and  $\delta \bm{u}^{(0)}$ is the incremental displacement vector. For the pressurized state, we have $\bm{x}_b^{(0)}=r_0\bm{e}_R+\lambda_zX\bm{e}_X$. For the plane-strain vibration modes, we set the components of $\delta \bm{u}^{(0)}$ to be
\begin{equation}\label{eq:6.2} 
\delta u^{(0)}_\Theta=U \exp(i(n\Theta-\omega t)),\quad \delta u^{(0)}_X=V \exp (i(n\Theta-\omega t)),\quad \delta u^{(0)}_R=W \exp(i(n\Theta-\omega t)),
\end{equation} 
where $(U, V, W)$ are constants, and $\omega$ is the angular
frequency and $n$ is the circumferential mode number.  Substituting the above two equations into the shell equations in cylindrical polar coordinates \eqref{eq:circular1}-\eqref{eq:circular3} and linearizing, one has three linear algebraic equations for $(U, V, W)$ in the form:
\begin{align}\label{eq:mMM}
\begin{pmatrix}
m_{11} & 0 & m_{13}\\
0 & m_{22} & 0\\
m_{31} & 0 & m_{33}
\end{pmatrix}\begin{pmatrix}
U\\
V\\
W
\end{pmatrix}
=\begin{pmatrix}
0\\
0\\
0
\end{pmatrix},
\end{align}
where the coefficients $m_{11}$, etc. are related to $n$, $\omega$ and the known quantities in the base state, whose expressions are omitted.  For the existence of nontrivial solutions, we need the determinant of the coefficient matrix to be zero, which leads to $D_1D_2=0$ with $D_{1}=m_{22}$ and $D_2=m_{11}m_{33}-m_{13}m_{31}$. We note that this equation gives a relation between the frequency and the material parameters of an artery; in particular, it may be used to determine the material parameters of an artery, if the technology is available to measure its vibration frequency.  The equation $D_1=0$ represents a purely axial motion with the only (incremental) displacement component $\delta u^{(0)}_X$ that is also independent of $X$, which is thus called the {\it axial mode}. The equation $D_2=0$
corresponds to the $X$-independent coupled motions with both circumferential and radial  displacements but without axial  displacements, which are called the {\it circumferential-radial mode} and {\it radial-circumferential mode} respectively. This way of naming is according to their displacement components when $n$ approaches zero. Precisely, when $n=0$, the circumferential-radial mode has the circumferential displacement only and the radial-circumferential mode has the radial displacement only. Now, we examine the effects of the axial stretch, pressure and fibre angle on the frequencies for different mode numbers $n$ (with the same  material and geometric parameters in the previous section). The numerical results will be displayed in terms of the non-dimensional frequency $\omega^*:={\omega 2h}/{\sqrt{c/\rho}}$.

We first investigate how the axial pre-stretch affects the  frequencies  of the plane-strain vibration modes of the pressurized artery. For fixed $P=1$ (kPa) and three different values of the axial pre-stretch $\lambda_z=1, 1.3, 1.6$, the frequencies  of the plane-strain vibration modes are shown in Table \ref{tab:b}.  The circumferential-radial mode with $n=1$ is not shown in the table, as it  represents a rigid body translation and thus has zero frequency, and the same reason applies to the axial mode with $n=0$; the circumferential-radial mode with $n=0$ is not shown because the frequency is not a real number.  It is seen that the frequencies of all vibration modes increase with the axial pre-stretch and the mode number expect the radial-circumferential mode, whose frequencies does not always increase with the axial pre-stretch.
\begin{table}[!h]
	\centering
	\caption{The frequencies of the plane-strain vibration modes at different axial pre-stretches (a) Axial mode (b) Circumferential-radial mode (c) Radial-circumferential mode }
		\begin{tabular}{c c c c c c}
		\hline 
		$\lambda_z$	 & $\omega^*, n=1$ & $\omega^*, n=2$ &  $\omega^*, n=3$ \\
		\hline
		$1$  & $0.330$ & $0.661$ &  $0.991$\\
		$1.3$ & $0.399$ & $0.799$ &  $1.199$\\
		$1.6$  & $0.503$ & $1.005$ &  $1.508$\\
		\hline\\
	\end{tabular}(a)\qquad
	\begin{tabular}{c c c c c c}
		\hline 
		$\lambda_z$  & $\omega^*, n=2$ &  $\omega^*, n=3$ \\
		\hline
		$1$ &  $0.354$ &  $0.636$\\
		$1.3$   & $0.392$ &  $0.699$\\
		$1.6$  & $0.438$ &  $0.785$\\
		\hline\\
	\end{tabular}(b)
	\begin{tabular}{c c c c c c}
		\hline 
		$\lambda_z$	&$\omega^*, n=0$ & $\omega^*, n=1$ & $\omega^*, n=2$ &  $\omega^*, n=3$ \\
		\hline
		$1$  & $0.525$ & $0.806$ & $1.290$ &  $1.820$\\
		$1.3$&  $0.511$ & $0.799$ & $1.288$ &  $1.822$\\
		$1.6$ & $0.529$ & $0.837$ & $1.348$ &  $1.905$\\
		\hline
	\end{tabular}(c)

	\label{tab:b}
\end{table}

Next we turn to determine the influence of the pressure on the frequencies of the plane-strain vibration modes. For fixed $\lambda_z=1$ and three different values of the pressure $P=0, 1, 2$ (kPa), the frequencies of the plane-strain vibration modes are shown in Table \ref{tab:e}. It is observed that the frequencies of all vibration modes increase with the pressure and the mode number.
\begin{table}[!h]
	\centering
	\caption{The frequencies of the plane vibration modes at different pressures (a) Axial mode (b) Circumferential-radial mode (c) Radial-circumferential mode }
	\begin{tabular}{c c c c c c}
		\hline 
		$P$ 	 & $\omega^*,n=1$ & $\omega^*,n=2$ &  $\omega^*,n=3$ \\
		\hline
		$0$ & $0.242$ & $0.484$ &  $0.727$\\
		$1$  & $0.330$ & $0.661$ &  $0.991$\\
		$2$  & $0.410$ & $0.820$ &  $1.229$\\
		\hline\\
	\end{tabular}(a)\qquad
	\begin{tabular}{c c c c c c}
		\hline 
		$P$	 & $\omega^*, n=2$ &  $\omega^*,n=3$ \\
		\hline
		$0$ & $0.032$ &  $0.154$\\
		$1$ & $0.354$ &  $0.636$\\
		$2$ & $0.505$ &  $1.017$\\
		\hline\\
	\end{tabular}(b)
	\begin{tabular}{c c c c c c}
		\hline 
		$P$ & $\omega^*,n=0$ & $\omega^*,n=1$ & $\omega^*,n=2$ &  $\omega^*,n=3$ \\
		\hline
		$0$& $0.463$ & $0.644$ & $1.006$ &  $1.401$\\
		$1$ & $0.525$ & $0.806$ & $1.290$ &  $1.820$\\
		$2$ & $0.644$ &$0.969$ & $1.552$ &  $2.181$\\
		\hline
	\end{tabular}(c)
	
	\label{tab:e}
\end{table}

Finally, we check the effect of the fibre angle on the frequencies of the plane-strain vibration modes. For fixed $\lambda_z=1$ and $P=1$ (kPa) and three different values of the fibre angle $\varphi=29^\circ, 45^\circ, 62^\circ$, the frequencies of the plane-strain  vibration modes are shown in Table \ref{tab:g}. It is seen that the frequencies of all vibration modes increase with the mode number. In addition, among the three vibration modes, the frequencies of the circumferential-radial mode and radial-circumferential mode decrease with the fibre angle, while frequencies of the axial mode  does not always decrease with the fibre angle.

\begin{table}[!h]
	\centering
	\caption{The frequencies of the plane-strain vibration modes at different fibre angles (a) Axial mode (b) Circumferential-radial mode (c) Radial-circumferential mode}
	\begin{tabular}{c c c c c c}
		\hline 
		$\varphi$	 & $\omega^*, n=1$ & $\omega^*, n=2$ &  $\omega^*, n=3$ \\
		\hline
		$29^\circ$   & $0.330$ & $0.661$ &  $0.991$\\
		$45^\circ$ & $0.364$ & $0.728$ &  $1.093$\\
		$62^\circ$ & $0.350$ & $0.699$ &  $1.049$\\
		\hline\\
	\end{tabular}(a)\qquad
	\begin{tabular}{c c c c c c}
		\hline 
		$\varphi$  & $\omega^*,n=2$ &  $\omega^*,n=3$ \\
		\hline
		$29^\circ$ & $0.354$ &  $0.636$\\
		$45^\circ$   & $0.344$ &  $0.599$\\
		$62^\circ$  & $0.322$ &  $0.541$\\
		\hline\\
	\end{tabular}(b)
	\begin{tabular}{c c c c c c}
		\hline 
		$\varphi$	 & $\omega^*,n=0$ & $\omega^*,n=1$ & $\omega^*,n=2$ &  $\omega^*,n=3$ \\
		\hline
		$29^\circ$ & $0.525$ & $0.806$ & $1.290$ &  $1.820$\\
		$45^\circ$&  $0.410$ & $0.661$ & $1.068$ &  $1.510$\\
		$62^\circ$& $0.259$ & $0.479$ & $0.798$ &  $1.139$\\
		\hline
	\end{tabular}(c)
	\label{tab:g}
\end{table}

\section{Concluding Remarks}
A consistent {\it static} finite-strain shell theory is available in the literature (see \cite{li2019consistent}), which involves three shell constitutive relations (deducible from the 3D constitutive relation) and six boundary conditions at each edge point. This work first presents a consistent {\it dynamic} finite-strain shell theory for incompressible hyperelastic materials in parallel. Novel aspects of our current study include:  1. The derivation of the refined shell equations through elaborate calculations which single out the bending effect with only two shell constitutive relations. 2. Many insights can be deduced from the refined shell equations. 3. It is not an easy task to get the proper number and proper form of physically meaningful boundary conditions in a shell theory. Here,  by using the weak form of the shell equations and the variation of the 3D Lagrange functional, four shell boundary conditions at each edge point are derived. 4. The 2D shell virtual work principle is obtained. A major advantage of this new shell theory is that its derivation does not involve any {\it ad hoc} kinematic or scaling assumptions (as almost all the existing derived shell theories for incompressible hyperelastic materials do). Due to its consistency with the 3D formulation in an asymptotic sense, one does not need to worry about its reliability in predicting the behaviors of incompressible hyperelastic shells for various loading conditions. In contrast, for assumptions-based shell theories some defects are evident. For example, some such shell theories involve higher-order stress resultants, whose physical meanings are not clear, and one does not know how to impose the proper boundary conditions for them. Another example is the Donnell shell theory, for which the traction from the top and bottom surfaces is assumed to be imposed on the middle surface, and if the shear traction on the top and bottom surfaces has the equal magnitude and opposite sign, that shell theory does not work. Another simple example is that some shell theories use the assumption that the thickness does not change,  which is obviously not valid when a large tensile load is applied at the edge (e.g., large uniform extension of a tube). Due to the simplicity of some assumptions-based shell theories, if, for particular applications, experiences/intuitions indicate that the assumptions involved do not cause a big error, by all means, they can be used. So, at least in theory, there are two differences between the present shell theory and those assumptions-based ones: prediction reliability (or confidence level) and generality. This shell theory is also tested against a benchmark problem: the  extension and inflation of an arterial segment. Good agreement with the exact solution to a suitable asymptotic order gives a verification of this shell theory. As an application to a dynamic problem, the plane-strain vibrations in a pressurized artery is considered, and the results reveal the influences of the axial pre-stretch, pressure and fibre angle on the vibration frequencies, which may be useful for determining the artery parameters.

Due to the space limit, we only present one application. In subsequent works, we intend to develop a general incremental shell theory by linearizing the present shell theory around a known base state. Then, we shall study wave propagation in an infinitely-long pressurized  artery and vibrations in all mode types in a finitely-long pressurized artery with suitable edge conditions. Analytical and numerical studies based on this shell theory for determining some post-bifurcation behaviors of incompressible hyperelastic shells will be left for future investigations. 

%
%
%
%
%
%
%
%

\begin{appendices}
	
\renewcommand\thesection{\Alph{section}.} 

\section{Some omitted expressions and calculations}

{\bf 1. Remark 3.1:} The expressions of $\bm{F}^{(2)}$ and $\bm{S}^{(2)}$ are given by
\begin{align}
\bm{F}^{(2)}&=2(\nabla\bm{x}^{(0)})\bm{k}^2+2(\nabla\bm{x}^{(1)})\bm{k}+\nabla\bm{x}^{(2)}+\bm{x}^{(3)}\otimes\bm{n},\\
\bm{S}^{(2)} & = \overline{\mathcal{A}}^{1}[\bm{F}^{(2)}] +\overline{\mathcal{A}}^{2}[\bm{F}^{(1)},\bm{F}^{(1)}]-2p^{(1)}\mathcal{R}^{1}[\bm{F}^{(1)}]-p^{(2)}\bm{R}^{0},
\end{align}
where 
\begin{align}
&\bm{A}^{0}=\frac{\partial W}{\partial\bm{F}}\Big|_{\bm{F}=\bm{F}^{(0)}},\quad \bm{R}^{0}=\frac{\partial R}{\partial\bm{F}}\Big|_{\bm{F}=\bm{F}^{(0)}}=\det(\bm{F}^{(0)})\bm{F}^{(0)-1},\\
&\mathcal{R}^{i} =\frac{\partial^{i+1}R}{\partial\bm{F}^{i+1}}\Big|_{\bm{F}=\bm{F}^{(0)}},\quad \overline{\mathcal{A}}^{i}=\mathcal{A}^{i}|_{\bm{F}=\bm{F}^{(0)}}-p^{(0)}\mathcal{R}^{i}.
\end{align}

{\bf 2. Below Equation (3.14):} The expressions of $p^{(2)}$ and $\bm{x}^{(3)}$ are given by
\begin{align}
\begin{split}
&p^{(2)} = \frac{1}{\bm{g}\cdot\bm{B}^{-1}\bm{g}}\big(\bm{g}\cdot\bm{B}^{-1}\bm{f}_3-\bm{R}^{0}[2\nabla\bm{x}^{(0)}\bm{k}^2+2\nabla\bm{x}^{(1)}\bm{k}+\nabla\bm{x}^{(2)}] \\
&\qquad\quad - \mathcal{R}^{(1)}[\bm{F}^{(1)},\bm{F}^{(1)}]-\bm{g}\cdot \bm B^{-1}(\rho\ddot{\bm{x}}^{(1)})\big),
\end{split}\\
&\bm{x}^{(3)} =\bm{B}^{-1}(\rho\ddot{\bm{x}}^{(1)}+p^{(2)}\bm{g}-\bm{f}_3),
\end{align}
with the vector $\bm{f}_3$ being
\begin{align}
\begin{split}
\bm{f}_3=&\nabla\cdot\bm{S}^{(1)}+\big(\overline{\mathcal{A}}^{1}[2(\nabla\bm{x}^{(0)})\bm{k}^2+2(\nabla\bm{x}^{(1)})\bm{k}+\nabla\bm{x}^{(2)}]+\overline{\mathcal{A}}^{2}[\bm{F}^{(1)},\bm{F}^{(1)}]\\
&-2p^{(1)}\mathcal{R}^{1}[\bm{F}^{(-1)}]\big)^T\bm{n} +(\bm{k}\bm{g}^\alpha)\cdot \bm{S}^{(0)}_{,\alpha}+\bm{q}^{(1)}_{b}.
\end{split}
\end{align}

{\bf 3. Above Equation (4.2):} Since (3.37) and (3.38) are deduced from subtracting  the 2D divergence of  (3.31) multiplied by $\bm{1}$ from the left from  (3.16), and (3.16) is derived from $\bm{S}^T\bm{n}|_{Z=2h}=\bm{q}^+$ with substitutions of the field equation and the bottom traction condition which are treated as identities, we conclude that (after dropping relatively higher-order terms)
\begin{align}
\bm{A}_t+A_3\bm{n}=-(\bm{S}^{T}\bm{n}|_{Z=2h}-\bm{q}^{+})-\nabla\cdot(\bm{1}\bm{C})=0.
\end{align}
From this equality, it is not hard to see that the terms related to $\delta\bm{x}(\bm{r},2h)$ in (4.1)  cancel each other.

{\bf 4. Equations (4.4) and (4.5):} The calculations of twisting moment $T$ and bending moment $M$ are given as follows.
\begin{align}
\begin{split}
T&=\int_0^{2h}(\bm{x}-\bm{x}(\bm{r},h)\times \bm{S}^T\bm{N})\cdot\bm{N}_m\sqrt{g}_\tau\,dZ=\int_{0}^{2h}(\bm{N}_m\times (\bm{x}-\bm{x}(\bm{r},h)))\cdot\bm{S}^T\bm{N}\sqrt{g_\tau}\,dZ\\
&=\int_{0}^{2h} (\bm{N}_m\times ((Z-h)\bm{x}^{(1)}+\frac{1}{2}(Z^2-h^2)\bm{x}^{(2)}))\cdot ((1+Z(\bm{k}-2H\bm{1}))\bm{S})^T\bm{\nu}\,dZ+O(h^4)\\
&=\int_{0}^{2h} (\bm{\nu}\times((Z-h)\bm{x}^{(1)}+\frac{1}{2}(Z^2-h^2)\bm{x}^{(2)}))\cdot (\bm{S}^{(0)}+Z\bm{S}^{(1)})^T\bm{\nu}\,dZ+O(h^4,h^3k)\\
&=\frac{2}{3}h^3\bm{S}^{(1)T}[\bm{\nu},\bm{\nu}\times \bm{x}^{(1)}]+\frac{1}{3}h^3\bm{S}^{(0)T}[\bm{\nu},\bm{\nu}\times \bm{x}^{(2)}]+O(h^4,h^3k).
\end{split}
\end{align}
A similar calculation shows
\begin{align}
\begin{split}
M&=\int_0^{2h}(\bm{x}-\bm{x}(\bm{r},h)\times \bm{S}^T\bm{N})\cdot\bm{T}_m\sqrt{g}_\tau\,dZ=\int_{0}^{2h}(\bm{N}_m\times (\bm{x}-\bm{x}(\bm{r},h)))\cdot\bm{S}^T\bm{N}\sqrt{g_\tau}\,dZ\\
&=\frac{2}{3}h^3\bm{S}^{(1)T}[\bm{\nu},\bm{\tau}\times \bm{x}^{(1)}]+\frac{1}{3}h^3\bm{S}^{(0)T}[\bm{\nu},\bm{\tau}\times \bm{x}^{(2)}]+O(h^4,h^3k).
\end{split}
\end{align}

{\bf 5. Equation (4.6):} The rotation angle $\alpha_m$ of the middle surface is calculated as follows. Let $C$ denote the intersection curve of the middle surface and the $\bm{n}\bm{N}_m$ plane and let $\bm{V}$ denote the tangent vector of $C$ at an edge point. Since the curve $C$ lies in the middle surface which is perpendicular to $\bm{n}$ and the $\bm{n}\bm{N}_m$ plane which is perpendicular to $\bm{T}_m$, we see that $\bm{V}$ is perpendicular to both $\bm{n}$ and $\bm{T}_m$. Thus $\bm{V}$ is the same  direction as the unit outward normal vector $\bm{N}_m$. We may take $\bm{V}=\bm{N}_m$ as the concern is about the angle not the magnitude. 

The curve $C$ can be parameterized by its arc length variable $q$ and we then have $\bm{V}=\frac{d\bm{X}_m(q)}{dq}$, where $\bm{X}_m=\bm{r}+h\bm{n}$ denotes the position vector of a point on the middle surface. After deformation, the curve $C$ is deformed into the curve $c:\bm{x}_m(q)$ with $\bm{x}_m=\bm{x}(\bm{X}_m)$. From the chain rule, the tangent vector of the curve $c$  is given by
\begin{align}
\bm{v}=\frac{d\bm{x}_m(q)}{dq}=\frac{d\bm{x}_m}{d\bm{X}_m}\frac{d\bm{X}_m(q)}{dq}=\nabla_m\bm{x}_m[\bm{V}]=\nabla_m\bm{x}_m[\bm{N}_m],
\end{align}
where $\nabla_m=\frac{\partial }{\partial\theta^\alpha}\widehat{\bm{g}}^\alpha|_{Z=h}=\frac{\partial }{\partial\theta^\alpha}(1-h\bm{k})^{-1}\bm{g}^\alpha$ is the 2D gradient operator on the middle surface.
Since $\alpha_m$ is defined as the angle between $\bm{v}$ projected to the $\bm{n}\bm{N}_m$ plane and the vector $\bm{N}_m$, we have
\begin{align}\label{eq:1}
\tan(\alpha_m)=\frac{\bm{v}\cdot\bm{n}}{\bm{v}\cdot\bm{N}_m}=\frac{\nabla_m\bm{x}_m[\bm{N}_m]\cdot\bm{n}}{\nabla_m\bm{x}_m[\bm{N}_m]\cdot\bm{N}_m}.
\end{align}

From the definition of $\nabla_m$, we see that $\nabla_m\bm{x}_m=\nabla\bm{x}_m+O(hk)$. Then by noticing that $\bm{x}_m=\bm{X}_m+\bm{u}_m$ and $\nabla\bm{n}=-\bm{k}$, we have
\begin{align}
\nabla_m\bm{x}_m=\nabla\bm{x}_m+O(hk)=\nabla (\bm{r}+h\bm{n})+\nabla\bm{u}_m+O(hk)=\bm{1}+\nabla\bm{u}_m+O(hk).
\end{align}
From equation (2.4) (in the manuscript), we have $ \sqrt{g_\tau}\bm{N}_m=(1+h(\bm{k}-2H\bm{1}))\bm{\nu}$, which implies that $\bm{N}_m=\bm{\nu}+O(hk)$. From these relations, \eqref{eq:1} can be simplified as
\begin{align}
\tan(\alpha_m)=\frac{\nabla\bm{u}_m[\bm{N}_m]\cdot\bm{n}}{1+\nabla\bm{u}_m[\bm{N}_m,\bm{N}_m]}+O(hk)=\frac{\nabla\bm{u}_m[\bm{\nu}]\cdot\bm{n}}{1+\nabla\bm{u}_m[\bm{\nu},\bm{\nu}]}+O(hk)
\end{align}
Then using the following two equalities
\begin{align}
&\nabla\bm{u}_m[\bm{\nu}]\cdot \bm{n}=\bm{u}_{m,\nu}\cdot\bm{n}=(\bm{u}_m\cdot\bm{n})_{,\nu}-\bm{u}_m\cdot\bm{n}_{,\nu}=u_{m3,\nu}+\bm{u}_m\cdot \bm{k}\bm{\nu}=u_{m3,\nu}+O(k),\\
&\nabla\bm{u}_m=\nabla(\bm{u}_{mt}+u_3\bm{n})=\nabla\bm{u}_{mt}+\bm{n}\otimes\nabla u_{m3}-u_{m3}\bm{k}=\nabla\bm{u}_{mt}+\bm{n}\otimes\nabla u_{m3}+O(k),
\end{align}
we conclude that
\begin{align}
\tan(\alpha_m)=\frac{u_{m3,\nu}}{1+\bm{1}\nabla\bm{u}_{mt}[\bm{\nu},\bm{\nu}]}+O(k,hk)
\end{align}

{\bf 6. Below Equation (4.10):} The calculations of  $L_i$ ($i=1,2,3$) are shown as follows. For $L_1$, it is calculated by
\begin{align}
\begin{split}
L_1&=\int_{\partial\Omega_q}\int_0^{2h}\bm{S}^T\bm{N}\cdot\delta \bm{u}_{mt}\,da=\int_{\partial\Omega_q}\int_0^{2h}((\bm{1}+Z(\bm{k}-2H\bm{1}))\bm{S})^T\bm{\nu}\cdot\delta \bm{u}_{mt}\,dZds\\
&=2h\int_{\partial\Omega_q}\overline{\bm{S}}^T_t\bm{\nu}\cdot\delta\bm{u}_{mt}\,ds+O(h^3).
\end{split}
\end{align}

For $L_2$, it is a sum of three terms, which will be calculated separately. The first term of $L_2$ can be calculated in the same way as $L_1$ and we have
\begin{align}
\begin{split}
&\int_{\partial\Omega_q}(\int_0^{2h} \bm{S}^T\bm{N}\cdot\bm{n}\sqrt{g_\tau}\,dZ)\delta u_{m3}\,ds=
\int_{\partial\Omega_q}(\int_0^{2h}((\bm{1}+Z(\bm{k}-2H\bm{1}))\bm{S})^T\bm{\nu}\cdot\bm{n}\,dZ)\delta u_{m3}ds\\
=&2h\int_{\partial\Omega_q}(\overline{\bm{S}}+\frac{2}{3}h^2\bm{1}\bm{S}^{(2)})^T\bm{\nu}\cdot\bm{n}\delta u_{m3}\,ds+O(h^4,h^4k)=2h\int_{\partial\Omega_q}(\overline{\bm{S}}\bm{n}+\frac{2}{3}\bm{1}\bm{S}^{(2)}\bm{n})\cdot\bm{\nu}\delta u_{m3}\,ds+O(h^4,h^4k).
\end{split}
\end{align}
Using (3.31) (in the manuscript), the above equation can be written as
\begin{align}
\begin{split}
\int_{\partial\Omega_q}(\int_0^{2h} \bm{S}^T\bm{N}\cdot\bm{n}\sqrt{g_\tau}\,dZ)\delta u_{m3}\,ds=&2h\int_{\partial\Omega_q}(\overline{\bm{S}}\bm{n}-(1-2Hh)\bm{1}\bm{S}^{(0)T}-h\bm{1}\bm{S}^{(1)T}+\frac{2}{3}h^2(\bm{1}\bm{S}^{(2)}\bm{n}-\bm{1}\bm{S}^{(2)T}\bm{n})\\
&-\frac{1}{3}\bm{1}\bm{S}^{(2)T}\bm{n}+\bm{m}_t)\cdot\bm{\nu}\delta u_{m3}\,ds+O(h^4,h^3k).
\end{split}
\end{align}
Upon using (3.4) (in the manuscript) and dropping any term which is relatively $O(h^2)$ or $O(h)$ smaller than another term, we have
\begin{align}
\begin{split}
\int_{\partial\Omega_q}\int_0^{2h} \bm{S}^T\bm{N}\cdot\bm{n}\sqrt{g_\tau}\,dZ=&2h\int_{\partial\Omega_q}(\overline{\bm{S}}\bm{n}-\overline{\bm{S}^T}\bm{n}+\frac{1}{3}h^2(\bm{1}\nabla\cdot\bm{S}^{(1)}_t-\rho\ddot{\bm{x}}_t^{1}+\bm{q}_{bt}^{(1)})\\
&+\bm{m}_t-h\bm{k}\bm{q}^-_t)\cdot\bm{\nu}\delta u_{m3}\,ds+O(h^4,h^3k).
\end{split}
\end{align}
The second term of $L_2$ is
\begin{align}
\begin{split}
&-\int_{\partial\Omega_q} (\int_{0}^{2h} (Z-h)\bm{S}^T\bm{N}\cdot(\bm{\nu}\times \bm{x}^{(1)})\sqrt{g_\tau}\,dZ)_{,s}\delta u_{m3}\,ds\\ 
=&-\int_{\partial\Omega_q} (\int_{0}^{2h} (Z-h)((\bm{1}+Z(\bm{k}-2H\bm{1}))\bm{S})^T\bm{\nu}\cdot(\bm{\nu}\times \bm{x}^{(1)})\,dZ)_{,s}\delta u_{m3}\,ds\\
=&-\int_{\partial\Omega_q} (\int_{0}^{2h} (Z-h)(\bm{S}^{(0)}+Z\bm{S}^{(1)})^T\bm{\nu}\cdot(\bm{\nu}\times \bm{x}^{(1)})\,dZ){,s}\delta u_{m3}\,ds+O(h^4,h^3k)\\
=&-\frac{2}{3}h^3\int_{\partial\Omega_q} (\bm{S}^{(1)T}[\bm{\nu},\bm{\nu}\times \bm{x}^{(1)}])_{,s}\delta u_{m3}\,ds+O(h^4,h^3k).
\end{split}
\end{align}
Similarly, the third term of $L_2$ is
\begin{align}
\begin{split}
&-\int_{\partial\Omega_q} (\int_{0}^{2h} \frac{1}{2}(Z^2-h^2)\bm{S}^T\bm{N}\cdot(\bm{\nu}\times \bm{x}^{(2)})\sqrt{g_\tau}\,dZ)_{,s}\delta u_{m3}\,ds\\ 
=&-\int_{\partial\Omega_q} (\int_{0}^{2h} \frac{1}{2}(Z^2-h^2)\bm{S}^{(0)T}\bm{\nu}\cdot(\bm{\nu}\times \bm{x}^{(2)})\,dZ){,s}\delta u_{m3}\,ds+O(h^4,h^3k)\\
=&-\frac{1}{3}h^3\int_{\partial\Omega_q} (\bm{S}^{(0)T}[\bm{\nu},\bm{\nu}\times \bm{x}^{(2)}])_{,s}\delta u_{m3}\,ds+O(h^4,h^3k).
\end{split}
\end{align}
Putting these terms together, we see that $L_2$ is given by
\begin{align}
\begin{split}
L_2=&2h\int_{\partial\Omega_q}\big((\overline{\bm{S}}\bm{n}-\overline{\bm{S}^T}\bm{n}+\frac{1}{3}h^2(\bm{1}\nabla\cdot\bm{S}^{(1)}_t-\rho\ddot{\bm{x}}_t^{1}+\bm{q}_{bt}^{(1)}))\cdot\bm{\nu}-\frac{1}{3}h^2(\bm{S}^{(1)T}[\bm{\nu},\bm{\nu}\times \bm{x}^{(1)}])_{,s}\\
&-\frac{1}{6}h^2(\bm{S}^{(0)T}[\bm{\nu},\bm{\nu}\times \bm{x}^{(2)}])_{,s}+\bm{m}_t\cdot\bm{\nu}-h\bm{k}\bm{q}^-_t\cdot\bm{\nu}\big)\delta u_{m3}\,ds+O(h^4,h^3k).
\end{split}
\end{align}

Similar calculations can be done for $L_3$ and we have
\begin{align}
\begin{split}
L_3&=-\int_{\partial\Omega_q}\int_{0}^{2h} (Z-h)\bm{S}^T\bm{N}\cdot (\bm{\tau}\times\bm{x}^{(1)})\delta\alpha_m\,da-\int_{\partial\Omega_q}\int_{0}^{2h}\frac{1}{2}(Z^2-h^2)\bm{S}^T\bm{N}\cdot (\bm{\tau}\times\bm{x}^{(2)})\delta\alpha_m\,da\\
&=-\frac{2}{3}h^3\int_{\partial\Omega_q}\bm{S}^{(1)T}[\bm{\nu},\bm{\tau}\times\bm{x}^{(1)}]\delta\alpha_{m\star}
\,ds-\frac{1}{3}h^3\int_{\partial\Omega_q}\bm{S}^{(0)T}[\bm{\nu},\bm{\tau}\times\bm{x}^{(2)}]\delta\alpha_{m\star}
\,ds+O(h^4,h^3k).
\end{split}
\end{align}

{\bf 7. Equation (6.2):} The expressions of $m_{ij}$ for the artery described by parameters in Table 1 when $P=4.33$ (kPa) and $\lambda_z=1$ are given by
\begin{align}
\begin{split}
m_{11}&=7.18153 - 136.457 n^2 + 1.14686 \omega^2,\\
m_{13}&= n((120.22i+9.05504i)n^2-0.0842880i)\omega^2,\\
m_{22}&=- 22.2934 n^2 + 1.29818 \omega^2,\\
m_{31}&=n(-130.636i+4.5984in^2-0.131045i\omega^2),\\
m_{33}&=-116.312 - 0.274395 n^4 + 1.29337 \omega^2 + 
n^2 (-9.45146 + 0.00240825 \omega^2).
\end{split}
\end{align}

\end{appendices}

\end{document}